\documentclass[12pt, draftclsnofoot, peerreview, onecolumn]{IEEEtran}

\IEEEoverridecommandlockouts

\usepackage{cite}
\usepackage{amsmath,amssymb,amsfonts}

\usepackage[linesnumbered,ruled,vlined]{algorithm2e}
\usepackage{graphicx}
\usepackage{textcomp}
\usepackage{xcolor}
\usepackage{booktabs}
\usepackage{array}
\usepackage{amsthm}
\usepackage{float}
\usepackage{setspace}
\usepackage{slashed}
\usepackage{multirow}

\def\BibTeX{{\rm B\kern-.05em{\sc i\kern-.025em b}\kern-.08em
    T\kern-.1667em\lower.7ex\hbox{E}\kern-.125emX}}

\newtheoremstyle{noparent} 
    {3pt} 
    {3pt} 
    {\normalfont} 
    {1em} 
    {\bfseries\itshape} 
    {.} 
    { } 
    {\thmname{#1} \thmnumber{#2}\mdseries\thmnote{ (\hspace{-0.25pt}#3)}} 
\theoremstyle{noparent}

\newtheorem{theorem}{\textbf{Theorem}}
\newtheorem{lemma}[theorem]{\textbf{Lemma}}
   
\newtheorem{corollary}[theorem]{\textbf{Corollary}}
\newtheorem{remark}[theorem]{\textbf{Remark}}

\newtheorem{definition}{\textbf{{Definition}}}

\makeatletter

\newcommand{\Rmnum}[1]{\expandafter\@slowromancap\romannumeral #1@}
\makeatother

\makeatletter
\renewenvironment{proof}[1][\proofname]{\par
  \pushQED{\qed}
  \normalfont \topsep6\p@\@plus6\p@\relax
  \trivlist
  \item[\hskip\labelsep
        \itshape
    #1\@addpunct{:}]\ignorespaces
}{
  \popQED\endtrivlist\@endpefalse
}
\makeatother

\DeclareMathOperator\supp{supp}

\DeclareMathOperator\bin{bin}

\SetKwIF{If}{ElseIf}{Else}{If}{then}{Else if}{Else}{Endif}
\SetKwFor{For}{For}{do}{Endfor}
\SetKwFor{While}{While}{do}{Endw}
\SetKwFunction{convb}{convb}

\begin{document}

\title{
    Construction and Design of MPAC Codes
}

\author{Fangbo Yi, Zuoxin Cai, Zhongjun Yang, {\em Student Member, IEEE}, Li Chen, {\em Senior Member, IEEE}, Huazi Zhang, {\em Senior Member, IEEE}, Wenxin Liu, and Yuan Li

\thanks{This article was presented in part at the 2023 IEEE International Symposium on Information Theory (ISIT) [DOI:10.1109/ISIT54713.2023.10206642]. 

Fangbo Yi, Zuoxin Cai, Zhongjun Yang and Wenxin Liu are with the School of Electronics and Information Technology, Sun Yat-sen University, Guangzhou 510006, China
(e-mail:  yifb@mail2.sysu.edu.cn; caizx7@mail2.sysu.edu.cn; yangzhj59@mail2.sysu.edu.cn; liuwx6@mail2.sysu.edu.cn). 

Li Chen is with the School of Electronics and Information Technology, Sun Yat-sen University, Guangzhou 510006, China, and also with Guangdong Province Key Laboratory of Information Security Technology, Guangzhou, 510006, China
(e-mail:  chenli55@mail.sysu.edu.cn). 

Huazi Zhang and Yuan Li are with Hangzhou Research Center, Huawei Technologies Co.,
Ltd., Hangzhou 310052, China (e-mail: zhanghuazi@huawei.com; liyuan299@huawei.com).

}}

\maketitle

\begin{abstract}
This paper proposes modified polarization-adjusted convolutional (MPAC) codes 
and their hybrid decoding 
that achieves an improved performance-complexity tradeoff. 
For MPAC codes, 
only a subset of the information bits undergo the convolutional transform. 
The output is then combined with the remaining information bits for the inner polar transform. 
Correspondingly, the convolutionally transformed bits are recovered by Fano decoding, while the remaining information bits are recovered by the successive cancellation (SC) decoding, constituting the hybrid Fano-successive cancellation (HFSC) decoding.
The MPAC codes are further designed by the 
coset-wise analysis that characterizes the number of minimum weight codewords (MWCs). 
It is discovered that a partially convolutional transform can improve the codeword through utilizing the row combinations of the frozen set efficiently.
This property enables the MPAC codes to outperform their prototype polarization-adjusted convolutional (PAC) codes and cyclic redundancy check (CRC)-polar codes.
Furthermore, MPAC codes can be optimized by reducing the number of MWCs.
Our numerical results demonstrate that, with a similar decoding complexity budget, 
the MPAC codes offer competent decoding performance when compared with PAC codes using Fano decoding and CRC-polar codes using SC list (SCL) decoding.

\end{abstract}


\section{Introduction}

Polar codes have been proven to achieve capacity of
the binary input discrete memoryless channel (BI-DMC), 
under the prerequisites of approaching infinite codeword length and 
the successive cancellation (SC) decoding \cite{Arikan2009ChannelPolarizationMethod}.
However, for short-to-medium length polar codes, 
the SC decoding performances remain limited. 
A notable improvement can be achieved through the SC list (SCL) decoding \cite{Tal2015ListDecodingPolar}\cite{Niu2012CRCAidedDecoding}, 
which keeps $L$ distinct instances of the SC decoding paths. 
The SCL decoding can approach the maximum likelihood (ML) decoding performance with a sufficiently large list size $L$.
Further improvement can be obtained by concatenating polar codes with 
the cyclic redundancy check (CRC) codes \cite{Tal2015ListDecodingPolar}\cite{Niu2012CRCAidedDecoding}, 
the parity check (PC) codes \cite{Wang2016ParityCheckConcatenated}\cite{Zhang2018ParityCheckPolar} or 
the convolutional codes \cite{Fazeli2017ViterbiAidedSuccessive}\cite{Arikan2019sequentialdecodingchannel}.

The polarization-adjusted convolutional (PAC) codes\cite{Arikan2019sequentialdecodingchannel}
concatenate an outer rate-$1$ convolutional code with an inner polar code. 
It is shown that the normal approximation (NA) bound can be approached by PAC codes using Fano decoding\cite{Arikan2019sequentialdecodingchannel}. 
However, PAC codes are usually constructed through the Reed-Muller (RM) rate profiling \cite{Arikan2019sequentialdecodingchannel}, 
which yields limited code rates. 
Addressing this, 
the existing polar code rate profiling schemes \cite{Arikan2009ChannelPolarizationMethod}\cite{Mori2009PerformancePolarCodes, Tal2013HowConstructPolar, Trifonov2012EfficientDesignDecoding, He2017BetaExpansionTheoretical} 
can be utilized for selecting the reliable subchannels, 
while the RM-polar rate profiling \cite{Li2014RMPolarCodes}\cite{Rowshan2021PolarizationAdjustedConvolutional} 
can yield both a flexible rate choice and improved codeword weight distribution. 
By further considering the effect of convolutional transform, 
performances of PAC codes can be improved. 
By considering subchannel utilizations, Liu {\em et al.} \cite{Liu2023} proposed the weighted sum (WS) metric for a more flexible rate profiling. 
It generalizes the RM rate profiling to design PAC code for any desired rate.
Several rate profiling designs have also been proposed in order to optimize
the decoding error probability \cite{Moradi2021MonteCarloBased}\cite{Chiu2023DesignPolarCodes}, 
reducing the Fano decoding complexity \cite{Moradi2024PACCodeRate}\cite{Jiang2023ConstructionPACCodes}.
The minimum weight distribution (MWD) contains knowledge of the code's minimum Hamming distance and 
the number of its minimum weight codewords (MWCs). 
Moreover, by considering the transpose of the convolutional transform matrix \cite{Gu2023ImprovedConvolutionalPrecoder} and 
formation of MWCs \cite{Rowshan2023FormationMinweight}, 
the MWD of PAC codes can also be improved. 
It leads to an enhanced ML decoding performance for the codes.

Fano decoding \cite{Fano1963heuristicdiscussionprobabilistic} is an efficient
algorithm in terms of its required hardware implementation resources, including memory and computation resources.
However, when the received information is unreliable, 
Fano decoding may suffer from a high decoding latency and complexity. 
To reduce the Fano decoding complexity, Moradi {\em et al.} \cite{Moradi2021SequentialDecodingMetric} 
proposed the path metric function that is parameterized by the subchannel cutoff rates.
Meanwhile, Rowshan {\em et al.} \cite{Rowshan2021PolarizationAdjustedConvolutional} 
proposed several tree search strategies to reduce the decoding complexity.
They both enable the Fano decoding to achieve a better performance-complexity tradeoff.
To further curb worst-case decoding latency and complexity, Fano decoding can be performed with a computational threshold \cite{Moradi2021SequentialDecodingMetric, Moradi2020PerformanceComplexitySequential, Liu2021HybridDecodingCRC}.
But it inevitably leads to a decoding performance degradation.
So far, limited efforts have been deployed through redesigning PAC codes in achieving a better performance-complexity tradeoff.

It is known that a code's ML decoding performance is determined by its
weight distribution. 
Both polar codes and PAC codes can be designed through optimizing their weight distributions 
\cite{MuradAbdullah2023NewSearchPolarization, Gelincik2022PreservingMinimumDistance, Rowshan2023FormationMinweight, Gu2023ImprovedConvolutionalPrecoder, Xu2017Distancespectrumoptimized}.
With an exhaustive search, weight distribution of both polar codes and PAC codes 
can be computed, but only up to the length of $64$ bits \cite{MuradAbdullah2023NewSearchPolarization}\cite{Xu2017Distancespectrumoptimized}.
Utilizing the recursive structure of polar codes, 
the probabilistic methods \cite{Valipour2013ProbabilisticWeightDistribution}\cite{Zhang2017EnhancedProbabilisticComputation} 
can estimate the codes' weight distribution. 
With a similar approach, Li {\em et al.} \cite{Li2021WeightSpectrumPrea} analyzed the average weight distribution
of the pre-transformed polar codes which include PAC codes. 
Overall, it remains challenging to compute a code's entire weight distribution.
Consequently, the MWD becomes a predominant metric in asymptotic performance analysis, as it dictates the error probability floor at high signal-to-noise ratio (SNR) regimes\cite{ShuLin2004ErrorControlCoding}.
For a family of polar codes that are defined by the partial order property, 
Bardet {\em et al.} \cite{Bardet2016Algebraicpropertiespolar_conf} provided an explicit formula 
for computing the number of MWCs. 
For the same family of codes, 
Rowshan {\em et al.} \cite{Rowshan2023FormationMinweight} characterized the row combinations of 
the MWCs under the coset-wise framework using a set-theoretic approach. 
It can be further generalized for analyzing the MWD of PAC codes \cite{Rowshan2023MinimumWeightCodewords}. 
Besides the above mentioned theoretical analysis, 
the MWD of both polar codes and PAC codes can also be numerically estimated. 
Li {\em et al.} \cite{Li2012AdaptiveSuccessiveCancellation} proposed the 
SCL-based enumeration method to evaluate the number of MWCs. But it requires a huge memory and complexity since the decoding output list size $L$ needs to be sufficiently large.

This paper proposes the modified PAC (MPAC) codes and their hybrid Fano-SC (HFSC) decoding. This coding scheme was first proposed in our previous work of \cite{Cai2023ModifiedPACCodes}, which performs convolutional transform only for a portion of information bits. It limits the number of nodes that are allowed for backtracking, yielding a better performance-complexity tradeoff than the relevant coding schemes.
In this work, the MPAC code design is further proposed based on the coset-wise analysis of its MWD, leading to the 
approximated union bound (AUB)-optimal MPAC codes that have a better MWD than PAC codes and CRC-polar codes.
It also shows the effectiveness of the partially convolutional transform design for improving the MWD. 
Major contributions of this work are summarized as follows:
\begin{itemize}
\item The MPAC codes are proposed. 
In the coding scheme, only a subset of information bits undergo the convolutional transform. 
Its output bits are then concatenated with the remaining information bits for the inner polar transform.
\item The HFSC decoding is further proposed to decode information bits. It recovers the convolutionally transformed information bits and the remaining information bits through Fano decoding and SC decoding, respectively. 
Our numerical results show that with a budgeted complexity, the proposed scheme yields a better performance-complexity tradeoff than Fano decoding of PAC codes and SCL decoding of CRC-polar codes.
\item The coset-wise analysis of the MWD of MPAC codes is presented, 
which leads to design the AUB-optimal MPAC codes.
By considering the impact of frozen bits in MPAC codes, 
it characterizes the row combinations of the MWCs. 
By further validating generation of the characterized codewords,
the exact volume of MWCs in MPAC codes can be obtained. 
For a wide range of code parameters,
the designed MPAC codes yield a better MWD than PAC codes and CRC-polar codes.
\end{itemize}

The rest of this paper is organized as follows. 
Section \ref{sec_pre} provides the background knowledge on polar codes and PAC codes.  
Section \ref{MPAC_HFSC} proposes the MPAC codes and their HFSC decoding.
Section \ref{sec4} analyzes the MWD of MPAC codes in the coset-wise framework.
Section \ref{AUBoptimal_MPAC_Codes} presents the design of AUB-optimal MPAC codes and their numerical results.
Finally, Section \ref{sec7} concludes this paper. 
Ahead of the rest sections, we first state our notation fashion as follows.

Let $\mathbb{F}_2=\{0,1\}$ and $\oplus$ denote the binary field and its addition operator, respectively.
We also let $\oplus$ denote the addition operator between the binary vectors.
Let $[l,u]$ denote the set $\{l, l+1, \ldots, u\}$, where $l<u$.
For a set $\mathcal{A}\subseteq[0,N-1]$, its cardinality and complementary set
are denoted by $|\mathcal{A}|$ and $\mathcal{A}^{c}$, respectively.
Further let $\underline{a}_i^j$, where $i<j$, denote the vector $(a_i, a_{i+1}, \ldots, a_j)$.
Given a vector $\underline{a}_{0}^{N-1}$ and a set $\mathcal{A} \subseteq [0,N-1]$,
we use $\underline{a}_{\mathcal{A}}$ to denote its subvector $(a_i \mid i\in\mathcal{A})$.
Note when $\mathcal{A}=[0,N-1]$, $\underline{a}_{0}^{N-1}=\underline{a}_{\mathcal{A}}$.
The binary representation of an integer $i\in [0, 2^n-1]$ is defined as $\bin(i)=i_{n-1}\ldots i_{1}i_{0}$,
where $i_{n-1}$ is the most significant bit and $i=\sum_{b=0}^{n-1}i_b 2^{b}$.
The support of a vector $\underline{a}_0^{N-1}$ is the index set of its non-zero coordinates, i.e., 
$\supp(\underline{a}_0^{N-1})\triangleq \{i\in [0,N-1] \mid a_i \neq 0\}$. 
Let $\mathcal{S}_i$ denote the support of $\bin(i)$, i.e., 
$\mathcal{S}_i\triangleq\{b\in [0,n-1] \mid  i_b \neq 0\}$.
Furthermore, let $\mathcal{S}_j \setminus \mathcal{S}_i$ denote all elements in the support of $\bin(j)$ but are not in the
support of $\bin(i)$.
The Hamming weight of a vector $\underline{c}_0^{N-1}\in \mathbb{F}_2^N$ is denoted as
${\rm w}(\underline{c}_0^{N-1})\triangleq |\supp(\underline{c}_0^{N-1})|$.
Given two vectors $\underline{c}=(c_0, c_1, \ldots, c_{N-1}) \in \mathbb{F}_2^N$ and 
$\underline{c}'=(c'_0, c'_1, \ldots, c'_{N-1}) \in \mathbb{F}_2^N$, 
the Hamming distance between them is defined as
${\rm d}(\underline{c}, \underline{c}')=|\{i\in[0,N-1] \mid c_i \neq c'_i\}|$.

\section{Polar Codes and PAC Codes}
\label{sec_pre}
This section provides the prerequisites for this work, 
including polar codes and its SC decoding, PAC codes and its Fano decoding.

\subsection{Approximated Union Bound (AUB)}
\label{MWCs_of_Polar_Codes}

A $K$-dimensional subspace $\mathcal{C}$ of $\mathbb{F}_2^N$ is called an $(N, K, d_{\min})$ 
binary linear block code, where $N$ and $K$ are the length and dimension of the code,
respectively, and the minimum distance $d_{\min}$ of the code is defined as 
$$d_{\min} = \min_{\underline{c},\underline{c}'\in \mathcal{C}, \underline{c}\neq\underline{c}'}
{\rm d}(\underline{c}, \underline{c}').$$

For a binary input additive white Gaussian noise (BI-AWGN) channel, 
union bound \cite{ShuLin2004ErrorControlCoding} on the ML decoding frame error rate (FER),
denoted as $P_e$,
of a binary linear block code is
\begin{equation}
    P_e \leq \sum_{d=d_{\min }}^N {\rm A}_d Q\left(\sqrt{2 d \cdot R \cdot E_{\rm b} / N_0}\right), \\
    \label{Union bound}
\end{equation}
where ${\rm A}_d$ denotes the number of weight-$d$ codewords, 
$Q(y)=\frac{1}{\sqrt{2\pi}} \int_{y}^{\infty} e^{-\frac{\vartheta^2}{2}}\,{\rm d}\vartheta$ 
is the tail distribution function of the standard normal distribution, 
$R={K}/{N}$ is the code rate
and $E_{\rm b} / N_0$ is the SNR per information bit.   
The MWD is defined by both $d_{\min}$ and ${\rm A}_{d_{\min}}$.
As SNR increases, MWD becomes dominant for (\ref{Union bound}).
Hence, the union bound can be approximated by 
\begin{equation}
    P_e \approx {\rm A}_{d_{\min}} Q\left(\sqrt{2 d_{\min} \cdot R \cdot E_{\rm b} / N_0}\right). \\
    \label{AUB}
\end{equation}
The above AUB is often utilized to assess the operational performance of a designed binary linear block code.

\subsection{Polar Codes and SC Decoding}
Polar codes are founded on channel capacity polarization
which is the consequence of channel combining and splitting \cite{Arikan2009ChannelPolarizationMethod}. 
Given a BI-DMC $\mathrm{W}: \mathcal{X} \rightarrow \mathcal{Y}$, 
with $\mathcal{X} \in \mathbb{F}_2$ and $\mathcal{Y} \in \mathbb{R}$,
a set of $N$ polarized subchannels 
{$\mathrm{W}_{N}^{(i)}: \mathcal{X} \rightarrow \mathcal{Y}^{N} \times \mathcal{X}^{i-1}$} can be obtained through the channel polarization process, where $N=2^{n}$, $n \in \mathbb{N}^{+}$ and $i \in [0,N-1]$.
Let $I(\mathrm{W})$ and $I(\mathrm{W}_{N}^{(i)})$ denote the symmetric capacity of
channel $\mathrm{W}$ and subchannel $\mathrm{W}_{N}^{(i)}$, respectively.
{Channel polarization results in most of the
subchannels $\mathrm{W}_{N}^{(i)}$ becoming either noiseless with $I(\mathrm{W}_{N}^{(i)}) \rightarrow 1$, 
or heavily noisy with $I(\mathrm{W}_{N}^{(i)}) \rightarrow 0$. }

An $(N, K)$ polar code is constructed by the rate profiling
and polar transform $\mathbf{G}_{\mathrm{p}}=\mathbf{F}^{\otimes n}$, 
where $\mathbf{F}=((1,0),(1,1))^{T} \in \mathbb{F}_{2}^{2 \times 2}$ 
is the kernel matrix and $\otimes$ denotes the Kronecker product.
Rate profiling with parameters $(N,K,\mathcal{I})$ embeds the information vector
$\underline{m}_{0}^{K-1} \in \mathbb{F}_{2}^{K}$ into a carrier vector 
$\underline{u}_{0}^{N-1} \in \mathbb{F}_{2}^{N}$.
It is defined by the information set $\mathcal{I} \subseteq [0,N-1]$ 
and the frozen set $\mathcal{I}^c=[0, N-1]\setminus\mathcal{I}$,
such that
$\underline{u}_{\mathcal{I}}=\underline{m}_{0}^{K-1}$ and $\underline{u}_{\mathcal{I}^c}=\underline{0}$.
After rate profiling, the codeword $\underline{c}_{0}^{N-1} \in \mathbb{F}_{2}^{N}$ can be generated by
polar transform as $\underline{c}_{0}^{N-1}=\underline{u}_{0}^{N-1} \mathbf{G}_{\mathrm{p}}$.
By denoting the polar transform matrix $\mathbf{G}_{\mathrm{p}}$ as
$\mathbf{G}_{\mathrm{p}}=[\mathbf{g}_0^T, \mathbf{g}_1^T, \ldots, \mathbf{g}_{N-1}^T]^T$,
the polar codeword can be defined as 
\begin{equation}
    \underline{u}_{0}^{N-1}\mathbf{G}_{\mathrm{p}}=u_{0}\mathbf{g}_{0}\oplus u_{1}\mathbf{g}_{1}\oplus \cdots \oplus u_{N-1}\mathbf{g}_{N-1}. \\
    \label{linear_combination_of_rows}
\end{equation}
With rate profiling that sets $\underline{u}_{\mathcal{I}^{c}}=\underline{0}$,
$\{\mathbf{g}_i \mid i\in\mathcal{I} \}$ forms a basis of the polar codebook, denoted as $\mathcal{C}(\mathcal{I})$.

Polar codes are decoded by
the SC algorithm in which 
the information bits are estimated successively 
based on their decision log-likelihood ratios (LLRs).
Let us assume that codeword $\underline{c}_{0}^{N-1}$ is transmitted through a noisy channel using binary phase shift keying (BPSK) modulation.
Let $\underline{y}_{0}^{N-1}=\left(y_{0}, y_{1}, \ldots, y_{N-1}\right) \in \mathbb{R}^{N}$
and $\underline{\hat{u}}_{0}^{N-1} = \left(\hat{u}_{0},\hat{u}_{1}, \ldots, \hat{u}_{N-1}\right) \in \mathbb{F}_{2}^{N}$  
denote the received symbol vector and the estimated carrier vector, respectively.
Over the SC decoding trellis, the decision LLR of the estimated bit $\hat{u}_{i}$ is defined as 
\begin{equation}
    L_{N}^{(i)}=
    \ln \frac{P(\underline{y}_{0}^{N-1}, \underline{\hat{u}}_{0}^{i-1} \mid \hat{u}_{i}=0)}{P(\underline{y}_{0}^{N-1}, \underline{\hat{u}}_{0}^{i-1} \mid \hat{u}_{i}=1)},
    \label{desision_LLRs}
\end{equation}
where $P(\underline{y}_{0}^{N-1}, \underline{\hat{u}}_{0}^{i-1} \mid \hat{u}_{i})$ is the transition
probability of $\mathrm{W}_{N}^{(i)}$. Estimation $\hat{u}_i$ can be successively made based on the decision LLR of (\ref{desision_LLRs}).
For $i\in \mathcal{I}^c$, $\hat{u}_{i}=0$ as they are frozen.
For $i\in \mathcal{I}$, $u_{i}$ is an information bit 
which can be estimated based on a bitwise ML decision rule as
\begin{equation}
    \hat{u}_{i}=\Upsilon(L_{N}^{(i)})=
    \begin{cases}
    0, & \text { if } L_{N}^{(i)} \geq 0\ ; \\
    1, & \text { otherwise}.
    \end{cases}
    \label{bit_decision}
\end{equation}
The SC decoding complexity is measured as the number of LLR computations, i.e., $N\log_2{N}$.

\subsection{PAC Codes and Fano Decoding}
\label{subsec_PAC_Codes_and_Fano_Decoding}

\vspace{-0.25cm}
\begin{figure}[htbp]
	\centering
	
    \includegraphics[scale=0.40]{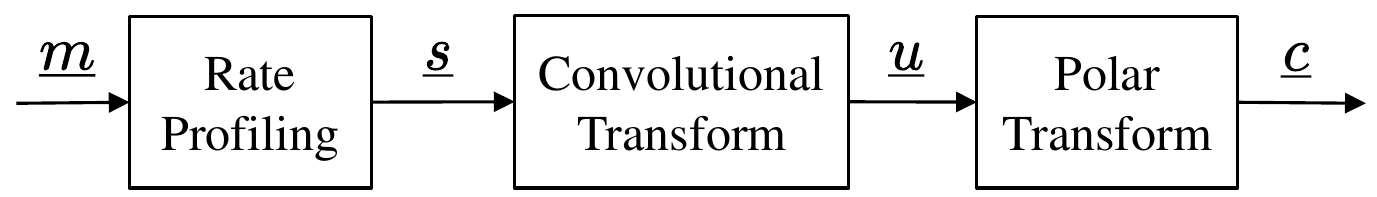}
    \vspace{-0.15cm}
	\caption{Block diagram of PAC codes.}
	\label{fig_PAC_encoding}
\end{figure}
A PAC code concatenates an outer rate-$1$ convolutional transform with an inner polar transform, 
as shown in Fig. \ref{fig_PAC_encoding}.
This concatenation protects the message bits in the inner code through utilizing their correlation provided by the convolutional transform, yielding an improved decoding performance.

Let $\underline{s}_{0}^{N-1} \in \mathbb{F}_{2}^{N}$ and $\underline{u}_{0}^{N-1} \in \mathbb{F}_{2}^{N}$
denote the input and output vectors of the convolutional transform, respectively.
To encode an $(N,K)$ PAC code, information vector $\underline{m}_{1}^{K}$ is
first embedded into $\underline{s}_{0}^{N-1}$ under the rate profiling with 
parameters $(N,K,\mathcal{I})$, such that
$\underline{s}_{\mathcal{I}}=\underline{m}_{0}^{K-1}$ and $\underline{s}_{\mathcal{I}^c}=\underline{0}$.

$\underline{u}_{0}^{N-1}$ is further generated 
by the convolutional transform as 
\begin{equation}
    u_i=\sum_{j=0}^{t} g_j s_{i-j},
    \label{1_bit_convolutional}
\end{equation}
where $\underline{g}_{0}^{t}=\left(g_{0}, g_{1}, \ldots, g_{t}\right)\in \mathbb{F}_{2}^{t+1}$ 
are coefficients of the convolutional generator polynomial.
Let $\underline{S}=(S[1], S[2], \cdots, S[t])$ denote state of the convolutional shift registers.
Algorithm \ref{algorithm_conv1b} (\texttt{conv1b($\cdot$)}) describes the generation of 
a convolutional output bit $u$ based on the current state $S$ and the input bit $v$.
Equivalently, the convolutional transform can be represented in 
an upper-triangular Toeplitz matrix $\mathbf{G}_{\mathrm{c}}\in \mathbb{F}_{2}^{N \times N}$,
whose rows are formed by shifting the vector $\underline{g}_{0}^{t}$.
Hence, $\underline{u}_{0}^{N-1}=\underline{s}_{0}^{N-1} \mathbf{G}_{\mathrm{c}}$.
Finally, the PAC codeword $\underline{c}_{0}^{N-1}$ is further generated by
$\underline{c}_{0}^{N-1}=\underline{u}_{0}^{N-1} \mathbf{G}_{\mathrm{p}}$.
Therefore, convolutional coded bits $\underline{u}_{0}^{N-1}$ are transmitted
through the polarized subchannels.

\begin{algorithm}[t]
    \label{algorithm_conv1b}
    \caption{Generation of a single convolutional output bit, \texttt{conv1b($\cdot$)}}
    \KwIn{$v$, $S$, $\underline{g}_{0}^{t}$;}
    \KwOut{$u$, $S$;}
    {\bf Initialize} $u=v\cdot g_0$;\\
    \For(\tcp*[f]{$|S|$ previous inputs}){ $j=1\to |S|$}{
        \If{$g_j=1$}{
            $u=u \oplus S[j]$;
        }
    }
    $S=(v,S[1], \cdots, S[t-1])$;\tcp*[f]{update shift register state}\\
    {\bf Return} $[u, S]$\;
\end{algorithm}

Fano decoding was first adopted to decode PAC codes, in which 
the constraints of the convolutional transform is imposed \cite{Arikan2019sequentialdecodingchannel}.
It can be seen as a depth-first search algorithm with the path metrics serving as the reliability indicators.
In this paper, we use the optimal path metric function \cite{Moradi2021SequentialDecodingMetric}, 
which can be computed layer-by-layer as
\begin{equation}
    M(\underline{\hat{u}}_{0}^{i})=M(\underline{\hat{u}}_{0}^{i-1}) + 1.0 + \log _{2} P(\hat{u}_{i} \mid \underline{y}_{0}^{N-1}, \underline{\hat{u}}_{0}^{i-1})\\
    -E_{0}(1, \mathrm{W}_{N}^{(i)}),
    \label{Fano_metric}
\end{equation}
where $P(\hat{u}_{i} \mid \underline{y}_{0}^{N-1}, \underline{\hat{u}}_{0}^{i-1})$ and 
$E_{0}(1, \mathrm{W}_{N}^{(i)})$ are the \textit{a posteriori} probability of $\hat{u}_{i}$
and the cutoff rate of subchannel $\mathrm{W}_{N}^{(i)}$, respectively.
For the BI-AWGN channel, $E_{0}(1, \mathrm{W}_{N}^{(i)})$ can be 
obtained by \cite{Gallager1968InformationTheoryReliable}
\begin{equation}
        E_{0}(1, \mathrm{W}_{N}^{(i)})=
        1-\log_2(1+\exp(-1/(2(\sigma_N^{(i)})^2))),
        \label{cutoff_rate}
\end{equation}
where $(\sigma_N^{(i)})^2$ is the simulated noise variance of $\mathrm{W}_{N}^{(i)}$. It can be estimated via the Gaussian approximation (GA) \cite{Trifonov2012EfficientDesignDecoding}.
By adjusting threshold $T$ dynamically with a step size $\Delta$ \cite{Fano1963heuristicdiscussionprobabilistic}, Fano decoding compares $M(\underline{\hat{u}}_{0}^{i})$ in controlling the decoding moves as:

\textbf{1)} If $\{M(\underline{\hat{u}}_{0}^{i})\}_{\rm{max}} \ge T $, 
move forward to path $\{\underline{\hat{u}}_{0}^{i}\}_{\rm{max}} $;

\textbf{2)} If $\{M(\underline{\hat{u}}_{0}^{i})\}_{\rm{max}}<T$, 
move backward to the root node of the tree, 
or the ancestor node $\hat{u}_{i'}$ which satisfies
$\{M(\underline{\hat{u}}_{0}^{i'})\}_{\rm{min}} > T $, where $i' \leq i$, and the path 
$\{\underline{\hat{u}}_{0}^{i'}\}_{\rm{min}}$ has not been visited.

At the beginning of the decoding, 
state of the shift registers and the path metric are initialized
as $\underline{S}=\underline{0}$ and $M\left(\underline{\hat{u}}_{0}^{-1}\right)=0$, respectively. 
Assuming that bit $\hat{u}_i$ ($i \in \mathcal{I}$) has been explored by Fano decoding and 
the current state of shift registers is $\underline{S}$,
the information bit $\hat{s}_i$ can be determined by the convolutional constraint as 
\begin{equation}
    \hat{s}_{i}=\lambda(\hat{u}_i, \underline{S})=
    \begin{cases}
    1, & \text { if } \texttt{conv1b}(1, \underline{S}, \underline{g}_{0}^{t}) = [\hat{u}_i, \underline{S}^{\prime}]\ ; \\
    0, & \text { otherwise},
    \end{cases}
    \label{conv_bit_decision}
\end{equation} 
where $\underline{S}^{\prime}$ means the next shift registers corresponding to the output estimation $\hat{u}_i$.
Furthermore, the next state $\underline{S}$ is updated by $\underline{S}^{\prime}$.
Once the decoding reaches a leaf of the tree, 
the complete estimations of $\underline{\hat{u}}_{0}^{N-1}$
and $\underline{\hat{s}}_{0}^{N-1}$ are obtained.

\section{MPAC Codes and HFSC Decoding}
\label{MPAC_HFSC}
This section introduces the MPAC codes and their HFSC decoding,
which can yield a good performance-complexity tradeoff.
Construction of the MPAC encoding includes two rate profilings which are defined by the proposed improved RM-polar (iRMP) orderings and the Gaussian approximation (GA), respectively.
The HFSC decoding integrates Fano decoding and SC decoding
to recover the information bits.

\subsection{Construction of MPAC Codes}
\label{Construction_of_MPAC_Codes}
\begin{figure}[htbp]
	\centering
	
    \includegraphics[scale=0.4]{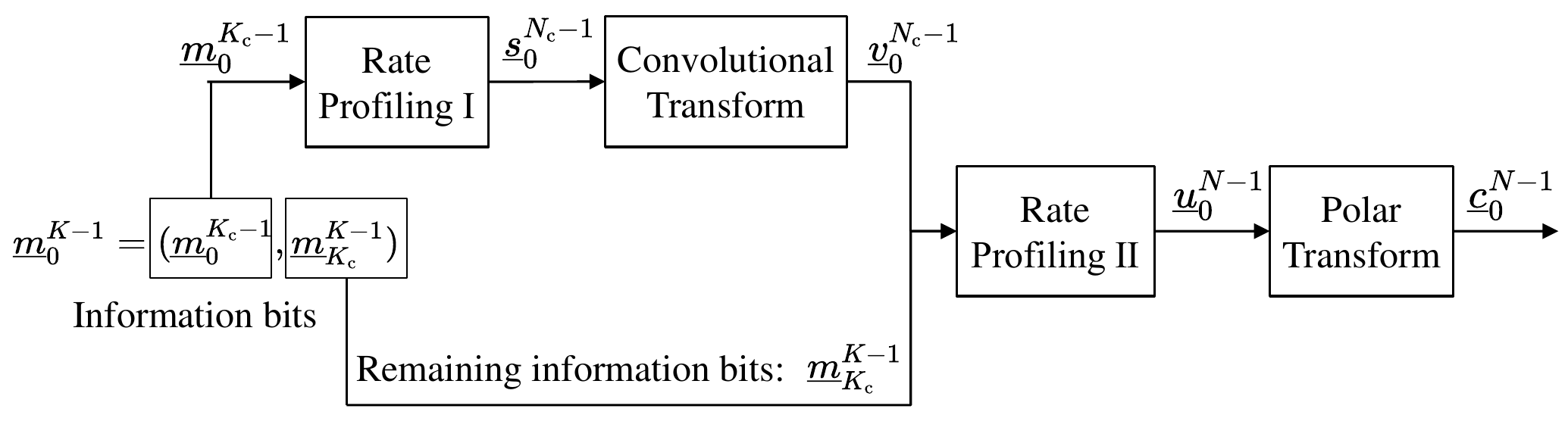}
    
	\caption{Block diagram of the $(N, K)$-$(N_{\rm c},K_{\rm c})$ MPAC codes.}
	\label{fig_MPAC_encoding}
\end{figure}

For an $(N, K)$ MPAC code, it needs to be further specified by the 
\textit{convolutional parameters} $(N_{\rm c},K_{\rm c})$, 
where $N_{\rm c}$ is the dimension of the convolutional transform as 
$\mathbf{G}^{\prime}_{\rm{c}} \in \mathbb{F}_{2}^{N_{\rm c} \times N_{\rm c}}$ and
$K_{\rm c}$ is the number of information bits that undergo the transform.
Fig. \ref{fig_MPAC_encoding} shows the encoding block diagram of an
$(N, K)$-$(N_{\rm c},K_{\rm c})$ MPAC code. 
The convolutional parameters should satisfy 
\begin{equation}
    K-K_{\rm c}+N_{\rm c}\leq N.
    \label{parameter_constraint}
\end{equation}
The equality holds when the MPAC code becomes the $(N, K)$ PAC code,
i.e., when $K_{\rm c}=K$ and $N_{\rm c}=N$. Hence, the MPAC codes can be considered as a more general case of PAC codes where the convolutional transform is performed on a portion of the information bits.

To encode an $(N,K)$-$(N_{\rm c},K_{\rm c})$ MPAC code, 
the information bits $\underline{m}_{0}^{K-1}$ is partitioned into 
$\underline{m}_{0}^{K-1}=(\underline{m}_{0}^{K_{\mathrm c}-1}, \underline{m}_{K_{\mathrm c}}^{K-1})$,
where $\underline{m}_{0}^{K_{\mathrm c}-1}$ and $\underline{m}_{K_{\rm c}}^{K-1}$
denote the information bits that will undergo the convolutional transform and 
the remaining information bits, respectively.
Furthermore, rate profiling \Rmnum{1} and \Rmnum{2} are utilized
to construct the input vectors of the convolutional transform and 
the polar transform, denoted as $\underline{s}_{0}^{N_{\mathrm c}-1}$
and $\underline{u}_{0}^{N-1}$, respectively.
The information bits $\underline{m}_{0}^{K_{\mathrm c}-1}$
will first be embedded into the vector $\underline{s}_{0}^{N_{\mathrm c}-1}$
under the rate profiling \Rmnum{1} with parameter $(N_{\rm c},K_{\rm c},\mathcal{D})$,
where $\mathcal{D} \subseteq [0, N_{\rm c}]$ and $|\mathcal{D}|=K_{\rm c}$.
$\underline{s}_{0}^{N_{\mathrm c}-1}$ is constructed as
$\underline{s}_{0}^{N_{\rm c}-1}=(\underline{s}_{\mathcal{D}}, \underline{s}_{\mathcal{D}^c})$,
where $\underline{s}_{\mathcal{D}}=\underline{m}_{0}^{K_{\rm c}-1}$ and $\underline{s}_{\mathcal{D}^c}=\underline{0}$.
The convolutional codeword is further generated by 
$\underline{v}_{0}^{N_{\mathrm c}-1}=\underline{s}_{0}^{N_{\mathrm c}-1} \mathbf{G}'_{\mathrm{c}}$.
It will be concatenated with the remaining information bits $\underline{m}_{K_{\rm c}}^{K-1}$ to be 
transmitted through the polarized subchannels.
The subchannels are chosen for transmitting information bits based on their reliabilities.
Let $\mathcal{A} \subseteq [0, N-1]$ denote the index set of the 
$K-K_{\rm c}+N_{\rm c}$ most reliable subchannels.
Among set $\mathcal{A}$, the $N_{\rm c}$ least reliable subchannels
are indexed by set $\mathcal{P}$.
Consequently, under rate profiling \Rmnum{2}, the vector $\underline{u}_{0}^{N-1}$
is partitioned into 
\begin{equation}
    \underline{u}_{0}^{N-1} =(\underline{u}_{\mathcal{A}\setminus\mathcal{P}}, \underline{u}_{\mathcal{P}},\underline{u}_{\mathcal{A}^c}),
    \label{u_vector}
\end{equation}
where $\underline{u}_{\mathcal{A}\setminus \mathcal{P} }=\underline{m}_{K_{\rm c}}^{K-1}$, $\underline{u}_{\mathcal{P}}=\underline{v}_{0}^{N_{\rm c}-1}$
and $\underline{u}_{\mathcal{A}^c}=\underline{0}$ 
are the remaining information bits, 
the convolutional codeword and the frozen bits, respectively.
Finally, the MPAC codeword is generated by
$\underline{c}_{0}^{N-1}=\underline{u}_{0}^{N-1} \mathbf{G}_{\rm{p}}$, where again $\mathbf{G}_{\rm{p}}$ is the polar transform matrix.

With the above description, it can be seen that the two 
rate profilings that are specified by the index sets $\mathcal{A}$, $\mathcal{P}$ and $\mathcal{D}$,
and the convolutional parameters $(N_{\rm c},K_{\rm c})$ determine an MPAC code.
They also interplay in determining the MPAC codes' optimal and 
practical error-correction performances.  
In particular, rate profiling \Rmnum{2} assigns the convolutional output 
and the remaining information bits based on the polarized subchannels reliabilities, 
while rate profiling \Rmnum{1} is further designed to 
provide a good weight distribution of MPAC codes.
This effect will be revealed in Section \ref{sec4}.
To realize this design,
GA \cite{Trifonov2012EfficientDesignDecoding} and 
an iRMP ordering \cite{Rowshan2021PolarizationAdjustedConvolutional} are utilized.
Selection of the convolutional parameters $(N_{\rm c},K_{\rm c})$
will be further discussed in Sections \ref{optimize_HFSC_performance} \ref{AUBoptimal_MPAC_Codes}.

\begin{figure}[htbp]
	\centering
	
    \includegraphics[scale=0.45]{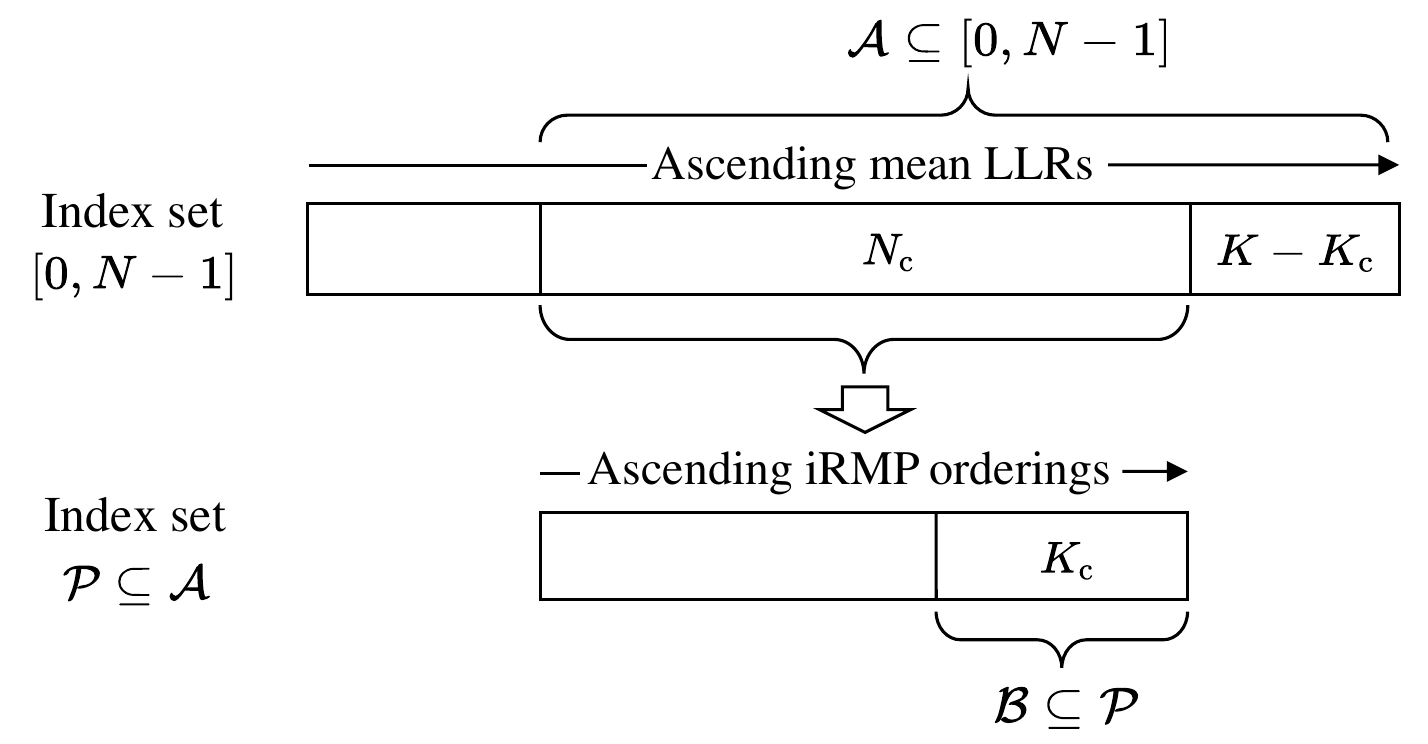}
    
	\caption{Construction of index sets.}
	\label{fig_MPAC_rate_profi}
\end{figure}

Fig. \ref{fig_MPAC_rate_profi} shows the construction of 
index sets in the two rate profilings.
It begins with the construction of index sets $\mathcal{A}$
and $\mathcal{P}$ of the rate profiling \Rmnum{2} using GA.
By assuming an all-zero codeword,
GA estimates the mean LLR value of all polarized subchannels $\mathrm{W}_{N}^{(i)}$,
which are denoted as ${\mathcal{L}}(\mathrm{W}_{N}^{(i)})$ and $i\in[0,N-1]$.
Note that the subchannel $\mathrm{W}_{N}^{(i)}$ with a greater ${\mathcal{L}}(\mathrm{W}_{N}^{(i)})$ 
is more reliable, and vice versa.
Hence, the $N$ subchannels can be ordered by their reliabilities,
yielding a refreshed subchannel index tuple $(j_{0}, j_{1}, \ldots, j_{N-1})_{\rm GA}$,
which implies 
$${\mathcal{L}}(\mathrm{W}_{N}^{(j_{0})})>{\mathcal{L}}(\mathrm{W}_{N}^{(j_{1})})>\cdots>{\mathcal{L}}(\mathrm{W}_{N}^{(j_{N-1})}).$$
Since $\mathcal{A}$ is the sorted index set of the 
$K-K_{\rm c}+N_{\rm c}$ most reliable subchannels, 
we have 
\begin{equation}
    \mathcal{A} = \{j_{0}, j_{1},\ldots, j_{K-K_{\rm c}+N_{\rm c}-1}\}.
    \label{A_set}
\end{equation}
Among $\mathcal{A}$, the index set of the $N_{\rm c}$ least reliable subchannels
is 
\begin{equation}
    \mathcal{P} = \{j_{K-K_{\rm c}}, j_{K-K_{\rm c}+1},\ldots, j_{K-K_{\rm c}+N_{\rm c}-1}\}.
    \label{P_set}
\end{equation}

When considering the design of rate profiling \Rmnum{1}, 
we also use $\mathcal{I}$ and $\mathcal{F}$ to 
denote the \textit{general information set} and \textit{general frozen set} 
w.r.t. the information set and the frozen set of the inner polar code, respectively.
For simplicity, subscripts $*_{\rm c}$ and $*_{\rm p}$ are used to denote those that will undergo 
the convolutional transform and those that will not, respectively. 
Hence, $\mathcal{I}=\mathcal{I}_{\rm c} \cup \mathcal{I}_{\rm p}$
and $\mathcal{F}=\mathcal{F}_{\rm c} \cup \mathcal{F}_{\rm p}$.
Consequently, rate profiling \Rmnum{2} constructs 
$\mathcal{I}_{\rm p}=\mathcal{A}\setminus \mathcal{P}$, 
$\mathcal{F}_{\rm p}=\mathcal{A}^c$ and 
$\mathcal{P}=\mathcal{I}_{\rm c} \cup \mathcal{F}_{\rm c}$ under the GA ordering, 
while rate profiling \Rmnum{1} only constructs $\mathcal{I}_{\rm c}$ and 
$\mathcal{F}_{\rm c}$.
Selecting the indices that indicate rows of higher weights in $\mathbf{G}_{\rm{p}}$ in forming $\mathcal{I}$ 
would result in a better weight distribution for the code \cite{Li2014RMPolarCodes}\cite{Rowshan2021PolarizationAdjustedConvolutional}.
For simplicity, let ${\rm r}(i)=2^{|\mathcal{S}_i|}={\rm w}(\mathbf{g}_i)$ denote the weight of row $\mathbf{g}_i$, where $i\in[0,N-1]$ and $\mathcal{S}_i$ denotes the support of $\bin(i)$.
To improve the weight distribution of an MPAC code, 
a feasible approach is to construct $\mathcal{I}_{\rm c}$
by selecting the indices with higher ${\rm r}(i)$ values from the set $\mathcal{P}$.
It can be realized by the iRMP rate profiling \cite{Rowshan2021PolarizationAdjustedConvolutional},
as it can often
result in a better weight distribution 
than the RM-polar rate profiling \cite{Li2014RMPolarCodes}.

\begin{definition}[iRMP Ordering]
    \label{iRMP_ordering}
    Given indices $i, j\in[0,N-1]$, we define iRMP ordering $(i,j)_{\rm iRMP}$
    if they satisfy either of the following conditions:
    \begin{itemize}
        \item ${\rm r}(i)>{\rm r}(j)$;
        \item ${\rm r}(i)={\rm r}(j)$ and ${\mathcal{L}}(\mathrm{W}_{N}^{(i)})\geq {\mathcal{L}}(\mathrm{W}_{N}^{(j)})$.
    \end{itemize}
    Note that the mean LLR of subchannel $\mathrm{W}_{N}^{(i)}$, i.e., ${\mathcal{L}}(\mathrm{W}_{N}^{(i)})$, 
    can be estimated by GA for $i \in [0,N-1]$.
\end{definition}

Since ${\mathcal{L}}(\mathrm{W}_{N}^{(i)}) \neq {\mathcal{L}}(\mathrm{W}_{N}^{(j)})$ for any $i,j\in[0,N-1]$ and $i\neq j$,
it is straightforward to verify that the iRMP ordering is a total order.
Note that any two indices belonging to $[0,N-1]$ are comparable under the iRMP ordering.

Under the iRMP ordering, the indices in set $\mathcal{P}$ are
ordered, yielding an $N_{\rm c}$-tuple 
$(j_{[0]},j_{[1]},\ldots,\\j_{[N_{\rm c}-1]})_{\rm iRMP}$
and $[\ *\ ]$ is a mapping of $[0,N_{\rm c}-1] \to [K-K_{\rm c},K-K_{\rm c}+N_{\rm c}-1]$.
The tuple implies that for any $a^\prime,b^\prime \in[0,N_{\rm c}-1]$ and 
$a^\prime < b^\prime$, we have $(j_{[a^\prime]},j_{[b^\prime]})_{\rm iRMP}$.
Indices with the $K_{\rm c}$ highest iRMP orderings are 
selected to construct set $\mathcal{B}$ as
\begin{equation}
    \mathcal{B} = \{j_{[0]}, j_{[1]},\ldots, j_{[K_{\rm c}-1]}\},
    \label{B_set}
\end{equation}
where $\mathcal{B} \subseteq \mathcal{P}$ and $|\mathcal{B}|=K_{\rm c}$.
Consequently, $\mathcal{I}_{\rm c}=\mathcal{B}$ and 
$\mathcal{F}_{\rm c}=\mathcal{P}\setminus\mathcal{B}$.

Index set $\mathcal{D}$ characterizes rate profiling \Rmnum{1} 
in disposing information bits $\underline{m}_{0}^{K_{\rm c}-1}$,
which is constructed based on sets $\mathcal{B}$ and $\mathcal{P}$ as follows.
Elements of $\mathcal{B}$ are first represented in an ascending order over a vector, 
yielding $\underline{b}_{0}^{K_{\rm c}-1}=\left(b_{0}, b_{1},\ldots, b_{K_{\rm c}-1}\right)$,
where $b_{0}< b_{1}<\cdots< b_{K_{\rm c}-1}$.
A mapping $\digamma_{\mathcal{B}}(*)$ is further defined as $\mathcal{B}=\{b_i|i\in[0,K_{\rm c}-1]\}\to[0, K_{\rm c}-1]$.
This implies that the index of the coordinate $b_i$ is determined by $i=\digamma_{\mathcal{B}}(b_i)$.
Similarly, mapping $\digamma_{\mathcal{P}}(*)$ is defined as $\mathcal{P}\to[0, N_{\rm c}-1]$.
Subsequently, index set $\mathcal{D}$ can be further constructed as 
\begin{equation}
    \mathcal{D}=\{k\in [0,N_{\rm c}-1] \mid k=\digamma_{\mathcal{P}}(b), b\in \mathcal{B}\},
    \label{D_set}
\end{equation}
where $|\mathcal{D}|=K_{\rm c}$.
Since $\mathcal{P}=\mathcal{I}_{\rm c} \cup \mathcal{F}_{\rm c}$ and
$\mathcal{B}=\mathcal{I}_{\rm c}$, mapping functions 
$\digamma_{\mathcal{I}_{\rm c} \cup \mathcal{F}_{\rm c}}(*)$ and 
$\digamma_{\mathcal{I}_{\rm c}}(*)$ are also defined, respectively. 
They are utilized to determine indices of the convolutional codeword 
$\underline{v}_{0}^{N_{\mathrm c}-1}$ and the information bits 
$\underline{m}_{0}^{K_{\rm c}-1}$ in the following HFSC decoding.
Note that the above mentioned index sets and mapping functions 
are computed once offline for a particular $(N,K)$-$(N_{\rm c},K_{\rm c})$ MPAC code.

\subsection{HFSC Decoding}
\label{The_HFSC_Decoding}
\begin{figure}[htbp]
	\centering
	
    \includegraphics[scale=0.40]{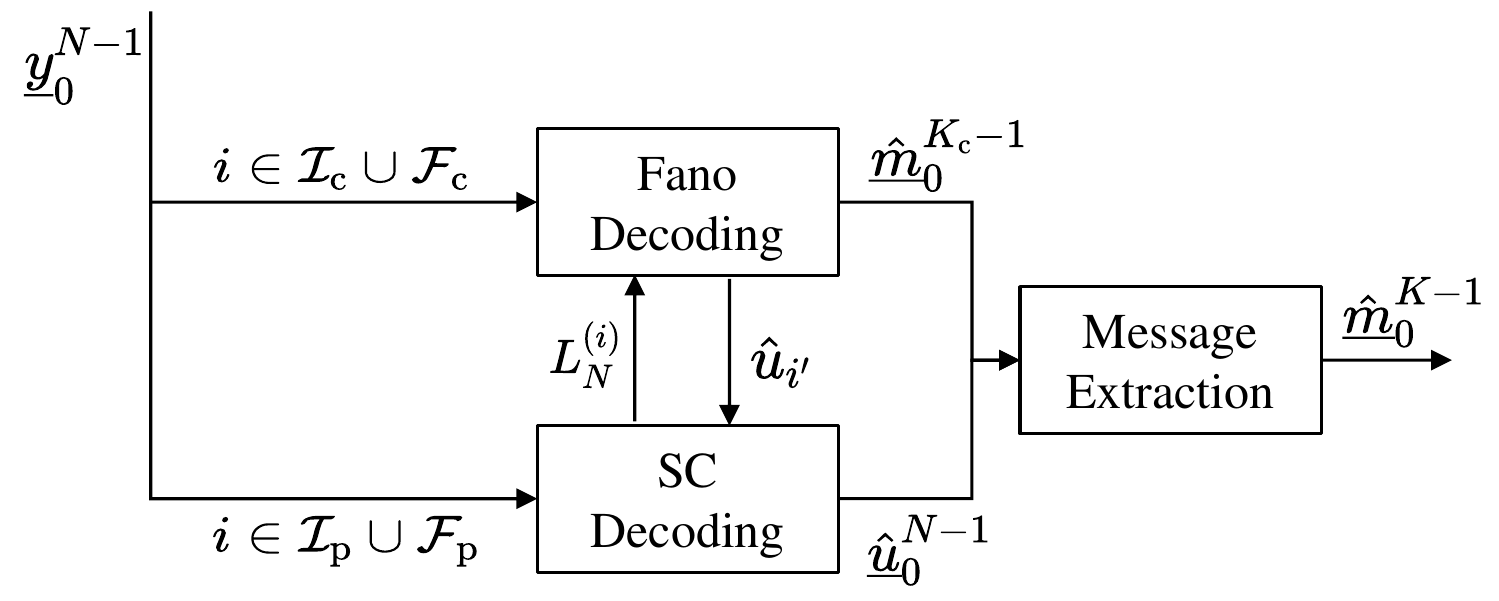}
    
	\caption{Block diagram of the HFSC decoding.}
	\label{fig_HFSC_decoding}
\end{figure}

Fig. \ref{fig_HFSC_decoding} shows block diagram of the HFSC decoding.
This is a successive information recovery, 
where the Fano decoding and the SC decoding are deployed for estimating the message bits, based on
the above mentioned index sets $\mathcal{I}_{\rm c}$, $\mathcal{F}_{\rm c}$, 
$\mathcal{I}_{\rm p}$, and $\mathcal{F}_{\rm p}$.
In an MPAC code, only partial information bits undergo the 
convolutional transform.  
The HFSC decoding can subsequently limit the number of nodes that allow backward computation in Fano decoding, and rationalize the decoding complexity.
It deploys the SC decoding to estimate the information bits $\underline{m}_{K_{\rm c}}^{K-1}$
that are transmitted through the most reliable subchannels, i.e., those
indexed by $\mathcal{I}_{\rm p}$.
Fano decoding will only be deployed to estimate the convolutional
codeword $\underline{v}_{0}^{N_{\rm c}-1}$ that are transmitted
through the subchannels indexed by 
$\mathcal{I}_{\rm c} \cup \mathcal{F}_{\rm c}$. 
Meanwhile, it further estimates the information bits $\underline{m}_{0}^{K_{\rm c}-1}$ 
that undergo the convolutional transform.
Note that during the decoding, 
the decision LLRs are computed by the SC decoding rule \cite{BalatsoukasStimming2015LLRBasedSuccessive}.
The HFSC decoding operates as follows.
\begin{itemize}
    \item For bits $u_i$ that are indexed by $i \in \mathcal{I}_{\rm c} \cup \mathcal{F}_{\rm c}$, 
        Fano decoding requests the SC rules for computing the decision LLRs $L_{N}^{(i)}$
        and further determines the path metrics $\{M(\underline{\hat{u}}_{0}^{i})\}_{\rm{max}} $ and 
        $\{M(\underline{\hat{u}}_{0}^{i})\}_{\rm{min}} $.
        Following a node visit as described in Section \ref{subsec_PAC_Codes_and_Fano_Decoding}, it
        estimates $\hat{u}_{i^{\prime}}$, where $i^{\prime}\leq i$. 
        This indicates that the decoder can potentially correct the previously decoded bits through backward computation.
        Furthermore, if $i' \in \mathcal{I}_{\rm c}$, the information
        bit $\hat{m}_{j}$ can be determined by the convolutional constraint
        of (\ref{conv_bit_decision}),
        where the index $j=\digamma_{\mathcal{I}_{\rm c}}(i')$.
        The Fano decoder feeds back 
        the estimated bits $\hat{u}_{i^{\prime}}$ to the SC decoder 
        and continues the decoding.
    \item For bits $u_i$ that are indexed by $i \in \mathcal{I}_{\rm p} \cup \mathcal{F}_{\rm p}$,
        the SC decoding estimates them based on 
        its decision LLR $L_{N}^{(i)}$ as in (\ref{bit_decision}).
        In particular, when $i\in \mathcal{F}_{\rm p}$, 
        it directly sets the frozen bits $\hat{u}_{i}=0$.
        It also updates the path metric $M(\underline{\hat{u}}_{0}^{i})$
        based on the estimated bit $\hat{u}_{i}$. 
\end{itemize}

Once $\hat{u}_{N-1}$ is estimated, 
or equivalently, when $i=N$, 
the information bits $\underline{\hat{m}}_{0}^{K-1}$ 
can be retrieved from both $\underline{\hat{m}}_{0}^{K_{\rm c}-1}$ 
and $\underline{\hat{u}}_{0}^{N-1}$.
Algorithm \ref{algorithm_HFSC} summarizes the HFSC decoding,
where the input index sets $\mathcal{I}_{\rm c}$, $\mathcal{F}_{\rm c}$, $\mathcal{I}_{\rm p}$ and $\mathcal{F}_{\rm p}$ are determined by the rate profilings \Rmnum{1} and \Rmnum{2}.

\begin{algorithm}[t]
    \label{algorithm_HFSC}
    \caption{The HFSC Decoding}
    \KwIn{$\underline{y}_{0}^{N-1}$, $\mathcal{I}_{\rm c}$, $\mathcal{F}_{\rm c}$, $\mathcal{I}_{\rm p}$, $\mathcal{F}_{\rm p}$;}
    \KwOut{$\underline{\hat{m}}_{0}^{K-1}$;}
    {\bf Initialize} $i=0$ and $j=0$;\\
    \While{$i \neq N$}{
        Compute decision LLR $L_{N}^{(i)}$ as in (\ref{desision_LLRs})\;
        \eIf(\tcp*[f]{Fano decoder}){$i \in \mathcal{I}_{\rm c} \cup \mathcal{F}_{\rm c}$}{
            Compute path metrics $M(\underline{\hat{u}}_{1}^{i})$ as in (\ref{Fano_metric})\;
            Perform a node visit of Fano decoding\;
            Estimate $\hat{u}_{i'}$ with $i' \leq i$ and let $i=i'+1$\;
            \If{$i'\in \mathcal{I}_{\rm c}$}{
            Let $j=\digamma_{\mathcal{I}_{\rm c}}(i')$ and determine $\hat{m}_{j}$ by convolutional constraint as in (\ref{conv_bit_decision})\;
            }
        }
        (\tcp*[f]{SC decoder})
        {
            \lIf{$i \in \mathcal{F}_{\rm p}$}{set $\hat{u}_i=0$}
            \If(\tcp*[f]{SC decoding}){$i \in \mathcal{I}_{\rm p}$}
            {Estimate $\hat{u}_i$ based on $L_{N}^{(i)}$ as in (\ref{bit_decision})}
            
            Compute $M(\underline{\hat{u}}_{0}^{i})$ as in (\ref{Fano_metric}) and let $i=i+1$\;
        }
    }
    Retrieve $\underline{\hat{m}}_{0}^{K-1}$ from $\underline{\hat{m}}_{0}^{K_{\rm c}-1}$ and $\underline{\hat{u}}_{0}^{N-1}$\;
    {\bf Return } $\underline{\hat{m}}_{0}^{K-1}$
\end{algorithm}
Note that the decoding computation threshold can also be utilized 
for the HFSC decoding to limit the decoding latency and complexity, 
which will be introduced in the following subsection.

\subsection{Simulation Results}
\label{optimize_HFSC_performance}

We first study the performance-complexity tradeoff of the proposed MPAC codes numerically. Its theoretical insights and improved design will be presented in Sections \ref{sec4}.
In this paper, all codes and decoding performances are evaluated over the AWGN channel 
using BPSK modulation.
The proposed schemes are compared with relevant coding schemes, 
including polar codes under SC decoding, PAC codes under Fano decoding, and
CRC-polar codes under SCL decoding.
Both the MPAC codes and PAC codes employ the
convolutional generator sequence $(1,0,1,1,0,1,1)$. 
MPAC codes are designed using GA at an SNR per codeword symbol of $E_{\rm s}/N_0=0$ dB,
while PAC codes are constructed using the RM rate profiling \cite{Arikan2019sequentialdecodingchannel}. 
For both the HFSC decoding and Fano decoding, 
the step size $\Delta=2$ which is numerically chosen in optimizing performance-complexity tradeoff \cite{Moradi2021SequentialDecodingMetric}.
Both the CRC-polar codes and polar
codes are designed using the reliability sequence in 5G standard \cite{20185GNRMultiplexing}. 
Given $N$ and $K$, an $(N,K)$-$(N_{\rm c}, K_{\rm c})$ MPAC code 
is alternatively denoted as MPAC-$(N_{\rm c}, K_{\rm c})$. 
The convolutional parameters $(N_{\rm c}, K_{\rm c})$ will be
determined by maximizing the HFSC decoding performance, which helps further verify the performance-complexity tradeoff advantage of the MPAC codes.

Decoding complexity is measured as the average number
of LLR computations required to decode a codeword.
A decoding LLR computation threshold $\Phi$ is introduced to bound the
decoding complexity and latency of both the HFSC decoding and Fano decoding. 
It is a multiplicity of the SC decoding complexity, 
i.e., $\Phi = \eta N \log_2 N$, where $\eta \in \mathbb{N} ^+$ is referred as the normalized complexity factor. 
For both the HFSC decoding and Fano decoding,
once $\Phi$ is reached, the SC decoding will be switched to in estimating the remaining information bits.
This prevents the decoding from lingering over the decoding tree, rationalizing the otherwise unlimited decoding computation.

\begin{figure}[!th]
	\centering
	\includegraphics[scale=0.47]{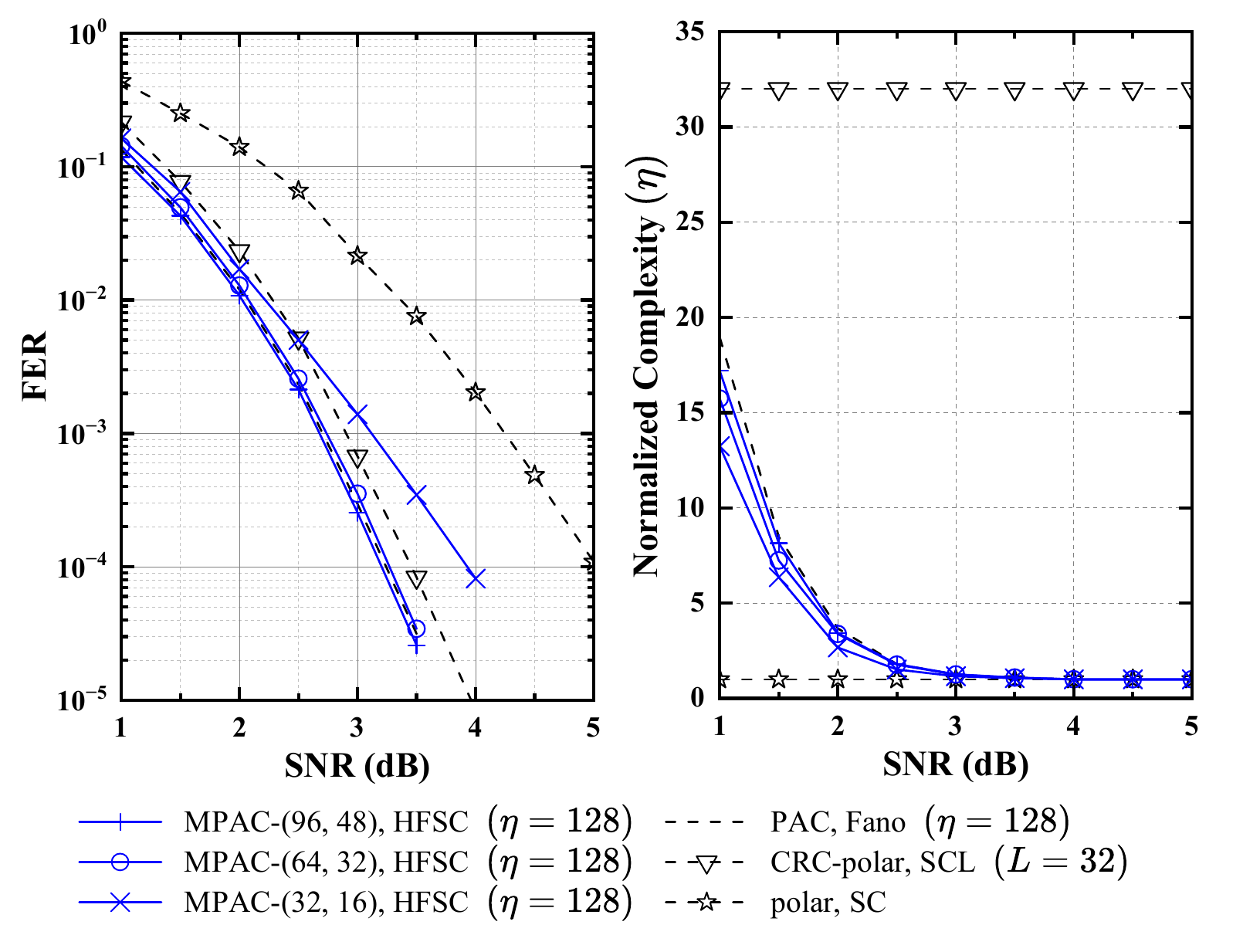}
	\caption{Performance comparison of MPAC codes, PAC codes, polar codes and CRC-polar codes, where $N = 128$ and $K = 64$.}
	\label{chap3_fig_simulation_128_64}
\end{figure}

Fig. \ref{chap3_fig_simulation_128_64} shows the decoding frame error rate (FER) and complexity performance of 
different coding schemes with $N=128$ and $K=64$. 
For yielding a good FER performance under the HFSC decoding,
optional parameters $(N_{\rm c}, K_{\rm c})=(96, 48)$, $(64, 32)$ and $(32, 16)$ are chosen for MPAC codes, which are obtained from our numerical study in \cite{Cai2023ModifiedPACCodes}.
A decoding computation threshold of $\eta=128$ is also provided for the HFSC decoding and Fano decoding as in \cite{Cai2023ModifiedPACCodes}. 
It can be seen that the MPAC-$(96,48)$ code can slightly outperform the $(128, 64)$ PAC code. 
Furthermore, when compared with the CRC-polar code, 
the MPAC-$(96,48)$ code achieves $0.25$ dB power gain at the FER of $10^{-4}$ 
with a smaller complexity.
It can also be seen that the MPAC-$(64,32)$ code performs nearly the same as 
the $(128, 64)$ PAC code, but with a lower decoding complexity.
Moreover, both the MPAC-$(96,48)$ code and MPAC-$(64,32)$ code can outperform the 
MPAC-$(32,16)$ code, validating the effectiveness of the convolutional 
parameters which are listed in \cite{Cai2023ModifiedPACCodes}.
It implys that a sufficient portion of information bits needs to undergo convolutional transform to achieve a good decoding performance.

\section{MWD Analysis}
\label{sec4}
This section analyzes the MWD of MPAC codes, 
including the formation of the MWCs and their coset-wise enumeration. 
Compared with the existing SCL-based enumeration \cite{Li2012AdaptiveSuccessiveCancellation},
it has a much lower computational complexity.
More importantly, this analysis provides a new
metric for designing MPAC codes, especially in determining the convolutional transform
parameters $(N_{\rm c}, K_{\rm c})$.        
We first revisit the formation of MWCs for polar codes.
By considering the impact of convolutional transform and frozen bits,
we present the general characterization of row combinations 
of the MWCs in MPAC codes together with the validation approach.
Based on this, an efficient algorithm is proposed to 
determine the number of MWCs.

\subsection{Formation of the MWCs for Polar Codes}
\label{Formation_of_the_MWCs_for_Polar_codes}
The set theoretical approach \cite{Rowshan2023FormationMinweight} 
in analyzing the MWD of polar codes is applied to show the intrinsic structure to 
the proposed MPAC codes.
By enumerating all the combinations, 
the formation and the number of MWCs can be obtained.
The approach can be used to analyze polar like codes with the partial order property, which will be defined as follows.

\begin{definition}[Partial Order\cite{Schurch2016}]
    \label{defi_partial_order}
    Given subchannel indices $i,j\in [0,N-1]$,
    the partial order $i \preceq j$ holds if 
    they satisfy one of the three conditions:
    \begin{itemize}
        \item $\mathcal{S}_{i} \subseteq \mathcal{S}_{j}$;
        \item $\mathcal{S}_{j}=(\mathcal{S}_{i}\setminus\{a\})\cup\{b\}$ for a pair of 
        $a\in \mathcal{S}_{i},b\notin\mathcal{S}_{i}$ and $a<b$;
        \item There exists an index $k \in\left[0,N-1\right]$ satisfying $i \preceq k$ and  $k \preceq j$.
    \end{itemize}
\end{definition}

Note that there may exist some pairs of indices
that neither of the index precedes the other under the partial order.
For this, the following Definition \ref{defi_POP} is needed.

\begin{definition}[Partial Order Property\cite{Rowshan2023FormationMinweight}]
    \label{defi_POP}
    A set $\mathcal{I}\subseteq\left[0,N-1\right]$ 
    is said to satisfy the partial order property if 
    none of the index pair $i\in\mathcal{I}$, 
    $i^{c}\in\mathcal{I}^{c}$ satisfies the partial order
    $i \preceq i^{c}$.
    Or equivalently, we have $i \npreceq i^{c}$ for any indices
    $i\in\mathcal{I}$ and $i^{c}\in\mathcal{I}^{c}$.
\end{definition}

With the above definition, if a set $\mathcal{I}$ 
satisfies the partial order property,
it can be found that all the indices in $\mathcal{I}$ 
compose a chain of partial orders. 
Moreover, for every pair of indices $i \in \mathcal{I}$ and $j \in[0, N-1]$, 
if $i \preceq j$, $j \in \mathcal{I}$.

Given a polar code $\mathcal{C}(\mathcal{I})$ with information set $\mathcal{I}$ that
satisfies the partial order property,
the analysis of its MWCs starts from dividing the code $\mathcal{C}(\mathcal{I})$ into  
$|\mathcal{I}|$ disjoint cosets. 
\begin{definition}[Polar Coset \cite{Rowshan2023FormationMinweight}]
    \label{definition_coset}
    Given an $N$-dimensional polar transform $\mathbf{G}_{\mathrm{p}}$ and
    an information set $\mathcal{I}\in[0,N-1]$, the coset is defined by
    \begin{equation}
        \mathcal{C}_i(\mathcal{I})=\left\{\mathbf{g}_i \oplus \bigoplus_{h\in\mathcal{H}}\mathbf{g}_h \mid \ \mathcal{H}\subseteq \mathcal{I}\setminus[0,i]\right\},
    \label{polar coset}
    \end{equation} 
    where the $i$-th row of $\mathbf{G}_{\mathrm{p}}$, i.e.,
    $\mathbf{g}_i$, is the coset leader.
\end{definition}
In other words, coset $\mathcal{C}_i(\mathcal{I})$ is the set
of codewords generated by $\mathcal{C}(\mathcal{I}\setminus [0,i-1])$,
where $\mathcal{C}(\mathcal{I}\setminus [0,i-1])$ is the subcode of 
polar code $\mathcal{C}(\mathcal{I})$.
The following Lemma \ref{corollary_weight_in_coset} indicates that
the Hamming weight of coset leader $\mathbf{g}_i$, where $i\in \mathcal{I}$ 
is a lower bound for the weight of any codeword in the coset $\mathcal{C}_i(\mathcal{I})$.

\begin{lemma}[\cite{Li2019PretransformedPolar,Rowshan2023FormationMinweight}]
    \label{corollary_weight_in_coset}
    For any $i\in[0,N-1]$ and $\mathcal{H}\subseteq[i+1,N-1]$, we have
    \begin{equation}
        {\rm w}(\mathbf{g}_i \oplus \bigoplus_{h\in\mathcal{H}}\mathbf{g}_h)\geq{\rm w}(\mathbf{g}_i).
        \label{w_coset_biger}
    \end{equation}
\end{lemma}

By comparing the row weights of all coset leaders, 
the minimum distance of the polar code $\mathcal{C}(\mathcal{I})$ can be obtained by 
\begin{equation}
    \label{minimum_weight}
    \begin{aligned}    
    d_{\min}&=\omega_{\min}\\&=\min(\{{{\rm w}(\mathbf{g}_i}) \mid i\in \mathcal{I}\}).
\end{aligned}
\end{equation}

Note that the weight of row $\mathbf{g}_i$
can be easily calculated by ${\rm w}(\mathbf{g}_i)=2^{|\mathcal{S}_i|}$, where $i\in[0,N-1]$.
Further let ${\rm A}_{\omega_{\min}}(\mathcal{I})$ and 
${\rm A}_{i, \omega_{\min}}(\mathcal{I})$ denote 
the number of MWCs in polar code $\mathcal{C}(\mathcal{I})$ and 
coset $\mathcal{C}_i(\mathcal{I})$, respectively.
Based on Lemma \ref{corollary_weight_in_coset},
we have ${\rm A}_{i, \omega_{\min}}(\mathcal{I})=0$
for every index $i\in\{j\in\mathcal{I} \mid {\rm w}({\bf g}_j)>\omega_{\min}\}$.
Consequently, only the cosets with 
their leaders' weight being equal to $\omega_{\min}$ need to be considered.
All linear combinations that form the MWCs in coset
$\mathcal{C}_i(\mathcal{I})$,
where $i\in \mathcal{I}$ and ${\rm w}({\bf g}_i)=\omega_{\min}$, 
can be explicitly characterized by the following
Lemma \ref{proposi_Polar_form}.

\begin{lemma}[\cite{Rowshan2023FormationMinweight,Bardet2016Algebraicpropertiespolar_conf}]
    \label{proposi_Polar_form}
    Given an information set $\mathcal{I} \subseteq [0,N-1]$ that satisfies the partial order property
    and coset leader index $i \in \mathcal{I}$ with ${\rm w}(\mathbf{g}_i)=\omega_{\min}$,
    the MWCs in polar coset $\mathcal{C}_i(\mathcal{I})$ can be formed 
    by rows as 
    \begin{equation}
        \label{polar_MWC_form}
        {\rm w}(\mathbf{g}_i \oplus \bigoplus_{j \in \mathcal{J}} \mathbf{g}_j \oplus \bigoplus_{m \in \mathcal{M}(\mathcal{J})} \mathbf{g}_m)=\omega_{\min},
    \end{equation}
    where the rows with index defined in sets $\mathcal{J}\subseteq \mathcal{K}_i$ and 
    $\mathcal{M}(\mathcal{J})$ are called the core rows and the balancing rows, respectively.
    Moreover, the sets $\mathcal{K}_i$ and $\mathcal{M}(\mathcal{J})$ can be constructed by
    the index of $i$ and the so called $\mathcal{M}$-construction, respectively.
\end{lemma}

For an index $i\in[0,N-1]$, the set $\mathcal{K}_{i}$ captures every index $j\in[i+1, N-1]$
that satisfies ${\rm w}(\mathbf{g}_j)\geq{\rm w}(\mathbf{g}_i\oplus\mathbf{g}_j)={\rm w}(\mathbf{g}_i)$.
Equivalently, set $\mathcal{K}_{i}$ is defined as 
\begin{equation}
\label{Ki_definition}
\mathcal{K}_i\triangleq \left\{ j\in \left[ i+1,N-1 \right]  \mid  |\mathcal{S}_j\setminus \mathcal{S}_i|=1 \right\}.
\end{equation} 
This set can be determined by applying the \textit{addition} operation and the \textit{left swap} operation on binary vector
$\bin(i)$ as in \cite{Rowshan2023FormationMinweight}.

Given a set $\mathcal{J}\subseteq\mathcal{K}_i$, 
the corresponding balancing set $\mathcal{M}(\mathcal{J})$ can be constructed using the
$\mathcal{M}$-construction \cite{Rowshan2023FormationMinweight}. 
However, the $\mathcal{M}$-construction is only conducted when all
the indices of set $\mathcal{J}$ are known.
It is more efficient to construct set $\mathcal{M}(\mathcal{J})$ through an 
ongoing approach \cite{Rowshan2022FastEnumerationMinimum} 
which enlists the indices of set $\mathcal{J}$ one-by-one.

The partial order property of 
information set $\mathcal{I}$ promises both 
sets $\mathcal{K}_i$ and $\mathcal{M}(\mathcal{J})$ are 
disjoint subsets of $\mathcal{I}\cap[i+1,N-1]$.
That says $\mathcal{K}_i\subseteq \mathcal{I}\cap[i+1,N-1]$,
$\mathcal{M}(\mathcal{J})\subseteq\mathcal{I}\cap[i+1,N-1]$
and $\mathcal{K}_i\cap\mathcal{M}(\mathcal{J})=\emptyset$.

As a result, the value of $\underline{u}_0^{N-1}$ can be controlled based on 
\begin{equation}
    \label{polar_u_setting}
    u_{x}=
    \begin{cases}
    1, & \text {if } x\in \{i\}\cup\mathcal{J}\cup\mathcal{M}(\mathcal{J});\\
    0, & \text {if } x\in [0,N-1]\setminus(\{i\}\cup\mathcal{J}\cup\mathcal{M}(\mathcal{J})).
    \end{cases}
\end{equation}
The corresponding MWCs can be further obtained by polar encoding as in (\ref{linear_combination_of_rows}).
Consequently, the number of MWCs in  
$\mathcal{C}(\mathcal{I})$ can be calculated as defined by 
the following Corollary \ref{corollary_MWC_number_polar}.
\begin{corollary}[\cite{Rowshan2023FormationMinweight}]
    \label{corollary_MWC_number_polar}
    Given an information set $\mathcal{I} \subseteq [0,N-1]$ that satisfies the partial order property, 
    the number of MWCs of $\mathcal{C}_i(\mathcal{I})$ is 
    \begin{equation}
    \label{polar_MWC_number}
    {\rm A}_{\omega_{\min}}(\mathcal{I})=\sum_{i \in \mathcal{I}:\ {\rm w}({\bf g}_i)=\omega_{\min}} 2^{\left|\mathcal{K}_{i}\right|}. 
    \end{equation} 
\end{corollary}

\subsection{Formation of the MWCs for MPAC Codes}
\label{sec5A}

\begin{figure}[htbp]
	\centering
    \includegraphics[scale=0.60]{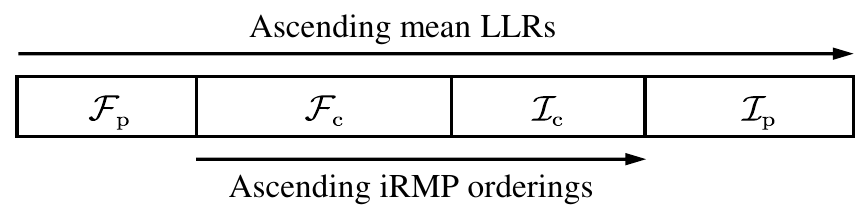}
	\caption{Index sets of $N$ subchannels under MPAC coding.}
	\label{fig_RP_MPAC}
\end{figure}

The above mentioned characterization of row combinations of 
MWCs for polar codes shows that it is important to define the index sets that correspond to 
the support of vector $\underline{u}_0^{N-1}$.
Therefore, in forming the MWCs of MPAC codes, the index sets $\mathcal{I}_{\rm c}$, 
$\mathcal{F}_{\rm c}$, $\mathcal{I}_{\rm p}$ and $\mathcal{F}_{\rm p}$. 
shall be characterized.
For our description convenience, they are also illustrated in Fig. \ref{fig_RP_MPAC}.

The general information set $\mathcal{I}=\mathcal{I}_{\rm c}\cup\mathcal{I}_{\rm p}$
and frozen set $\mathcal{F}=\mathcal{F}_{\rm c}\cup\mathcal{F}_{\rm p}$
are further obtained.
The rate profiling \Rmnum{1} ensures the subchannel indexes $i$ with higher ${{\rm w}(\mathbf{g}_i})$ are chosen, yielding a higher minimum row weights, i.e., $\min(\{{{\rm w}(\mathbf{g}_i}) \mid  i\in \mathcal{I}\}$, for MPAC codes.

It is worth mentioning that the method proposed for MWCs enumeration of MPAC code can be realized through checking the legality of every MWC enumeration.
As a result, their MWCs can be formed by (\ref{polar_MWC_form}) and further 
be enumerated by the row combination manner as (\ref{polar_u_setting}).
In other words, we need to set $u_x=1$ for every 
$x\in \{i\}\cup\mathcal{J}\cup\mathcal{M}(\mathcal{J})$.
However, it has been observed in \cite{Rowshan2023MinimumWeightCodewords}
that the outer convolutional transform with 
the frozen bits as inputs generates constraint on the values of some $u_x$.
In the proposed MPAC codes, 
$\underline{u}_{0}^{N-1}=(\underline{u}_{\mathcal{I}_{\rm c}\cup\mathcal{F}_{\rm c}},
\ \underline{u}_{\mathcal{I}_{\rm p}},\ \underline{u}_{\mathcal{F}_{\rm p}})$,
where the values of $\underline{u}_{\mathcal{I}_{\rm c}\cup\mathcal{F}_{\rm c}}$, 
$\underline{u}_{\mathcal{I}_{\rm p}}$ and $\underline{u}_{\mathcal{F}_{\rm p}}$ 
are the output vector of convolutional transform, information bits 
and frozen bits, respectively.
In the following, we will discuss the influence of convolutional transform and the frozen bits
on the value of $u_x$, which are characterized in three cases.

\textbf{\textit{Case \Rmnum{1}:}} When $x\in \mathcal{I}_{\rm c}\cup\mathcal{F}_{\rm c}$, 
$u_x$ is the convolutional output,
which is determined by $u_x=\bigoplus_{j=0}^{t} g_{j} s_{k-j}$, or equivalently, 
by applying \texttt{conv1b($\cdot$)}, where $k=\digamma_{\mathcal{I}_{\rm c} \cup \mathcal{F}_{\rm c}}(x)$.
Considering the constraint of $t$ previous convolutional inputs, 
we have $u_x=V \oplus s_{k}$,
where $V=\bigoplus_{j=1}^{t} g_{j} s_{k-j}$.
Since the value of $s_{k}$ is determined by either an information bit or a frozen bit, 
we consider the corresponding subcases of $x\in \mathcal{I}_{\rm c}$ and $x\in \mathcal{F}_{\rm c}$.

\textbf{\textit{\ Case \Rmnum{1}-A:}} When $x\in \mathcal{I}_{\rm c}$, $s_{k}=0$ or $1$.
Hence, we can obtain $u_x=1$ by choosing
$s_{k}$ appropriately such that 
\begin{equation}
    \label{conv_1b_chosen}
    s_{k}=
    \left\{
    \begin{array}{ll}
    1, & \text {if } V=0;\\
    0, & \text {if } V=1.
    \end{array}
    \right.
\end{equation}
In obtaining $u_x=0$, we can take the opposites of (\ref{conv_1b_chosen}).

\textbf{\textit{\ Case \Rmnum{1}-B:}} When $x\in \mathcal{F}_{\rm c}$, $s_{k}$ is 
determined by a frozen bit,
and let $s_{k}=0$.
Consequently, $u_x$ is only determined by
the previous input constraint, 
i.e., $u_x=V$.
If the previous inputs cause $V=1$, we have $u_x=1$.
Otherwise, $u_x=0$.

\textbf{\textit{Case \Rmnum{2}:}} When $x\in \mathcal{I}_{\rm p}$, 
$u_x$ is determined by an information bit,
$u_x=0$ or $1$.

\textbf{\textit{Case \Rmnum{3}:}} When $x\in \mathcal{F}_{\rm p}$,
$u_x$ is directly determined by a frozen bit and $u_x=0$.

It can be seen that in \textit{\ Case \Rmnum{1}-A} and \textit{\ Case \Rmnum{1}-B}, we have no control on the 
value of $u_x$.
Based on the setting of $\underline{u}_{0}^{N-1}$ in obtaining the MWCs as
in (\ref{polar_u_setting}), we focus on the value of $u_x$
that is only determined by convolutional operation, where $x\in (\mathcal{F}_{\rm c}\cup\mathcal{F}_{\rm p})\cap[i+1, N-1]$.
Since in \textbf{Case \Rmnum{3}} $u_x=0$,
we turn to focus on \textbf{Case \Rmnum{1}-B}.
Consequently, the index set $\tilde{\mathcal{F}_{\rm c}}$ is defined as 
\begin{equation}
    \label{F_c_piao}
\tilde{\mathcal{F}_{\rm c}}  \triangleq \left\{f \in \mathcal{F}_{\rm c} \cap [i+1, N-1] | u_f=1\right\}.
\end{equation}
The generation of $\tilde{\mathcal{F}_{\rm c}}$ implies that 
another term $\bigoplus_{f \in \tilde{\mathcal{F}_{\rm c}}} \mathbf{g}_f$ 
will be included in the row combination of (\ref{polar_MWC_form}).
A subset of $\tilde{\mathcal{F}_{\rm c}}$ 
is further defined as
\begin{equation}
\label{F_J}
\mathcal{F}_{\mathcal{J}}=\{f \in \tilde{\mathcal{F}_{\rm c}}  \mid  \left|\mathcal{S}_f \setminus \mathcal{S}_i\right|=1 \}.
\end{equation}

\begin{remark}
    \label{remark_7}
    The inclusion of the additional term $\bigoplus_{f \in \tilde{\mathcal{F}_{\rm c}}} \mathbf{g}_f$ 
    in the MWC row combination of (\ref{polar_MWC_form}) 
    does not reduce the minimum distance of an MPAC code.
    This is due to the fact that for any $f \in \tilde{\mathcal{F}_{\rm c}}$, 
    the enlisted row $\mathbf{g}_f$ is not the coset leader of any coset in the code.
    Based on Lemma \ref{corollary_weight_in_coset}, 
    since the general information set $\mathcal{I}=\mathcal{I}_{\rm c}\cup\mathcal{I}_{\rm p}$ remains unchanged,
    the minimum distance of the MPAC code is maintained as
    \begin{equation}
        \label{dmin_MPAC}
    \begin{aligned}        
        d_{\min}&=\omega_{\min}\\&=\min(\{{{\rm w}(\mathbf{g}_i}) \mid  i\in \mathcal{I}\ \text{and}\ \mathcal{I}=\mathcal{I}_{\rm c}\cup\mathcal{I}_{\rm p}\}).
    \end{aligned}
\end{equation}
\end{remark}

Please note that the above property does not always hold for all MPAC codes.
But in this work, we only consider the MPAC codes that yield (\ref{dmin_MPAC}).
Theorem \ref{theorem_MPAC_form} further characterizes the more
general row combination in obtaining the MWCs of an MPAC code.

\begin{theorem}
    \label{theorem_MPAC_form}
    
    Given an MPAC code that is defined by the subchannel index sets $\mathcal{I}_{\rm c}$, $\mathcal{F}_{\rm c}$, $\mathcal{I}_{\rm p}$ and $\mathcal{F}_{\rm p}$ where index $i$ is chosen such that ${\rm w}(\mathbf{g}_i)=\omega_{\min}$,
    the minimum weight codewords in coset $\mathcal{C}_i(\mathcal{I})$ can be further redefined based on (\ref{F_J}) as
    \begin{equation}
        \label{MPAC_MWC_form}
        {\rm w}(\mathbf{g}_i \oplus \bigoplus_{j \in \mathcal{J} \cup \mathcal{F}_\mathcal{J}} \mathbf{g}_j \oplus \bigoplus_{m \in \mathcal{M}(\mathcal{J} \cup \mathcal{F}_\mathcal{J})} \mathbf{g}_m )=\omega_{\rm min},
    \end{equation}
    where $\mathcal{J} \subseteq \mathcal{K}_i$ and $\mathcal{M}(\mathcal{J})$ 
    are constructed as in Lemma \ref{proposi_Polar_form}.
    The set $\mathcal{F}_{\mathcal{J}}$ can be represented based on (\ref{F_c_piao}) and (\ref{F_J}) as
    $$\mathcal{F}_{\mathcal{J}}=\left\{f \in \mathcal{F}_{\rm c}\cap[i+1,N-1] \mid u_f=1\ \text{and}\ |\mathcal{S}_f \setminus \mathcal{S}_i|=1\right\}.$$
\end{theorem}

\begin{proof}[\hspace{1.5em}Proof]
    It is a generalization of Lemma \ref{proposi_Polar_form}.    
    The main difference between MWC generation of polar codes and that of MPAC codes lies in the index set $\mathcal{F}_\mathcal{J}$.
    It is caused by the convolutional constraints and 
    the frozen bits as in \textbf{Case \Rmnum{1}-B}.
    Furthermore, since the indices of set $\mathcal{F}_\mathcal{J}$ have the same property 
    as those of the set $\mathcal{J}\subseteq\mathcal{K}_i$,
    we can replace the sets $\mathcal{J}$ and $\mathcal{M}(\mathcal{J})$ of (\ref{polar_MWC_form}) 
    by $\mathcal{J} \cup \mathcal{F}_\mathcal{J}$
    and $\mathcal{M}(\mathcal{J} \cup \mathcal{F}_\mathcal{J})$, respectively.
    This immediately leads to (\ref{MPAC_MWC_form}).
    In the following subsection \ref{sec:verification_MWCs}, we will propose an efficient approach to yield sets $\mathcal{J}$ and $\mathcal{F}_\mathcal{J}$, which can be enumerated one by one as in {\bf Algorithm \ref{algorithm_Ad_MPC}}
\end{proof}

\begin{figure}[htbp]
	\centering
	
    \includegraphics[scale=0.65]{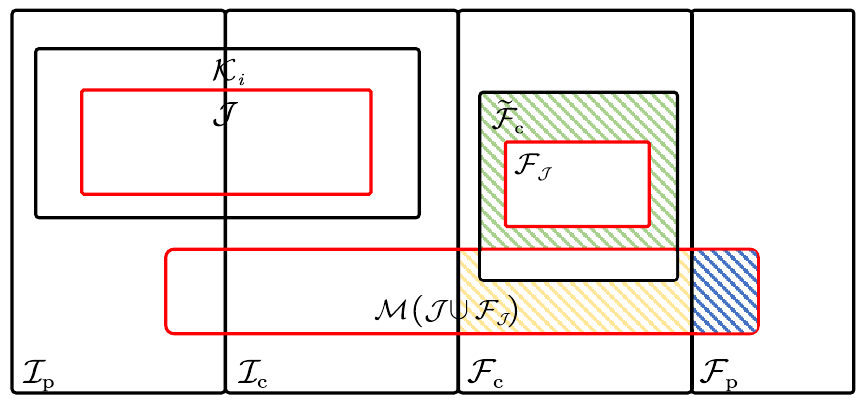}
	\caption{Venn diagram of the associated index sets for indices $[i+1,N-1]$.}
	\label{fig_venn_diagram}
\end{figure}

Fig. \ref{fig_venn_diagram} illustrates Venn diagram of the associated index sets in characterizing the MWCs of MPAC codes.
In particular, the index sets of rows that form the MWCs
are denoted by the red boxes.
We further combine these index sets into the set
\begin{equation}
    \label{MPAC_MWC_setting_Ui}
    \mathcal{U}_i(\mathcal{J})=\{i\} \cup (\mathcal{J} \cup \mathcal{F}_\mathcal{J}) \cup \mathcal{M}(\mathcal{J} \cup \mathcal{F}_\mathcal{J}),
\end{equation}
such that in obtaining the MWC, it should be 
\begin{equation}
    \label{MPAC_MWC_setting_u}
    u_x=
    \begin{cases}
    1 & \text { if } x \in \mathcal{U}_i(\mathcal{J}); \\
    0 & \text { if } x \in [0, N-1]\setminus\mathcal{U}_i(\mathcal{J}).
    \end{cases}
\end{equation}

Therefore, the number of MWCs of the proposed MPAC codes can be determined.
However, as discussed in \textbf{Case \Rmnum{1}-B} and \textbf{Case \Rmnum{3}}, for $f\in\mathcal{F}_{\rm c}\cup\mathcal{F}_{\rm p}$,
we have no control on the value of $u_f$.
If eq. (\ref{MPAC_MWC_setting_u}) cannot be ensured, the corresponding MWC cannot be formed.
The following subsection introduces the verification of the  MWCs for MPAC codes and determines their exact number.

\subsection{Verification of the MWCs for MPAC Codes}
\label{sec:verification_MWCs}

Based on Corollary \ref{corollary_MWC_number_polar}, 
the number of MWCs characterized in (\ref{MPAC_MWC_form})
can be obtained by enumerating all the possible 
index sets $\mathcal{U}_i(\mathcal{J})$ that are defined as in (\ref{MPAC_MWC_setting_Ui}).
Since the index set $\mathcal{M}(\mathcal{J} \cup \mathcal{F}_\mathcal{J})$ is constructed based on
set $\mathcal{J} \cup \mathcal{F}_\mathcal{J}$,
all possible index sets $\mathcal{J} \cup \mathcal{F}_\mathcal{J}$ need to be found\cite{Rowshan2023MinimumWeightCodewords}.

\begin{lemma}
    \label{lemma_MPAC_upper_bound}
    The number of all possible index sets  $\mathcal{J} \cup \mathcal{F}_\mathcal{J}$ is equal to $2^{\left|\mathcal{K}_{i}\right|}$.
\end{lemma}

\begin{proof}[\hspace{1.5em}Proof]
    The number of all possible index sets $\mathcal{J}\subseteq\mathcal{K}_i$ 
    is $2^{\left|\mathcal{K}_{i}\right|}$.
    Since set $\mathcal{F}_\mathcal{J}$ is generated based on the 
    convolutional outputs, it is related to the previous convolutional inputs.
    When obtaining the MWCs of an MPAC code, the convolutional inputs are 
    determined by (\ref{MPAC_MWC_setting_u}).
    In other words, $\mathcal{F}_\mathcal{J}$ depends on set $\mathcal{J}$.
    Therefore, the number of all possible sets $\mathcal{J} \cup \mathcal{F}_\mathcal{J}$ 
    is equal to that of all possible sets $\mathcal{J}$, 
    i.e., $2^{\left|\mathcal{K}_{i}\right|}$.
\end{proof}

Combining both \textbf{Case \Rmnum{1}-B} and \textbf{Case \Rmnum{3}} of Section \ref{sec5A},
$u_x=\bigoplus_{j=1}^{t} g_{j} s_{k-j}=V$ and
$k=\digamma_{\mathcal{I}_{\rm c} \cup \mathcal{F}_{\rm c}}(x)$ for $x\in \mathcal{F}_{\rm c}$ 
and $u_x=0$ for $x\in \mathcal{F}_{\rm p}$, respectively.
Consequently, for some $\mathcal{J}\subseteq\mathcal{K}_i$,
the value of $u_x$ and $x\in \mathcal{F}_{\rm c}\cup \mathcal{F}_{\rm p}$, cannot be chosen to ensure
$\supp({\underline{u}_0^{N-1}})=\mathcal{U}_i(\mathcal{J})$.
This results in some codewords that are formed as (\ref{MPAC_MWC_form}) 
cannot be generated by the MPAC encoding.
For better representation, let us further define
$$\mathcal{E}_{\mathcal{J}}^{1}=\{(\tilde{\mathcal{F}_{\rm c}}\setminus\mathcal{F}_{\mathcal{J}}) \setminus 
((\mathcal{F}_{\rm c}\cap[i+1,N-1])\cap\mathcal{M}(\mathcal{J} \cup \mathcal{F}_\mathcal{J}))\},$$
$$ \mathcal{E}_{\mathcal{J}}^{2}=\{((\mathcal{F}_{\rm c}\cap[i+1,N-1])\cap\mathcal{M}(\mathcal{J} \cup \mathcal{F}_\mathcal{J})) 
\setminus \tilde{\mathcal{F}_{\rm c}}\, $$
$$\mathcal{E}_{\mathcal{J}}^{3}=\{(\mathcal{F}_{\rm p}\cap[i+1,N-1])\cap\mathcal{M}(\mathcal{J} \cup \mathcal{F}_\mathcal{J})\}.$$ 

For a more illustrative presentation, the index sets 
$\mathcal{E}_{\mathcal{J}}^{1}$, $\mathcal{E}_{\mathcal{J}}^{2}$ and $\mathcal{E}_{\mathcal{J}}^{3}$ are 
marked by the green, yellow and blue shades
in Fig \ref{fig_venn_diagram}, respectively.
Considering the contradiction of $\supp({\underline{u}_0^{N-1}})\neq \mathcal{U}_i(\mathcal{J})$,
there are three criteria that can be utilized to validate the case
in which the characterized codewords cannot be generated.
For every $\mathcal{J}\subseteq\mathcal{K}_i$, when $x\in \mathcal{F}_{\rm c}\cup \mathcal{F}_{\rm p}$, the value of $u_x$ 
needs to be checked as follows.\\
\indent\textbf{Criterion 1 (C1)}: $u_x=1$, but $x\in \mathcal{E}_{\mathcal{J}}^{1}$;\\
\indent\textbf{Criterion 2 (C2)}: $u_x=0$, but $x\in \mathcal{E}_{\mathcal{J}}^{2}$;\\
\indent\textbf{Criterion 3 (C3)}: $u_x=0$, but $x\in \mathcal{E}_{\mathcal{J}}^{3}$.

In \textbf{C1}, since $x\in \mathcal{E}_{\mathcal{J}}^{1}$, $u_x$ should be set 
as $0$ such that (\ref{MPAC_MWC_setting_u}) is met. 
However, we get $u_x=1$ under the constraints of 
\textbf{Case \Rmnum{1}-B}, 
which implies the undesired row $\mathbf{g}_x$ has been included in the 
row combinations of (\ref{MPAC_MWC_form}).
It prevents formation of the corresponding MWC.
In \textbf{C2}, since $x\in \mathcal{E}_{\mathcal{J}}^{2}$, $u_x$ should be set as $1$.
However, under the same constraints in \textbf{Case \Rmnum{1}-B}, 
we get $u_x=0$, resulting in 
the desired row $\mathbf{g}_x$ being excluded from the row combinations of (\ref{MPAC_MWC_form}).
\textbf{C3} also checks the case in which the characterized MWCs
cannot be generated due to the absence of some specific rows.
Note that a missing row is caused by the constraint
of the corresponding frozen bit in \textbf{Case \Rmnum{3}}
and its index $x\in \mathcal{E}_{\mathcal{J}}^{3}$.
With the above three criteria, 
the exact number of the MWCs of an MPAC code is characterized as in 
Theorem \ref{theorem_MPAC_number}.

\begin{theorem}
    \label{theorem_MPAC_number}
    
    With the subchannel index sets $\mathcal{I}_{\rm c}$, $\mathcal{F}_{\rm c}$,
    $\mathcal{I}_{\rm p}$ and $\mathcal{F}_{\rm p}$ of MPAC codes,
    the number of MWCs in coset $\mathcal{C}_i(\mathcal{I})$ of the MPAC code is
    \begin{align}
        {\rm A}_{i,\omega_{\min}}^{\rm MPAC} = 2^{|\mathcal{K}_i|} - X,
    \end{align}
    where $X$ is the number of the MWCs that are characterized in Theorem \ref{theorem_MPAC_form},
    but cannot be formed. It is called the MWC number deficit.
\end{theorem}
\begin{proof}[\hspace{1.5em}Proof]
    Based on Lemma \ref{lemma_MPAC_upper_bound},
    the number of MWCs in $\mathcal{C}_i(\mathcal{I})$ of an MPAC code 
    has an upper bound of $2^{|\mathcal{K}_i|}$.
    The upper bound is achieved only if for all $\mathcal{J}\subseteq\mathcal{K}_i$,
    the corresponding vector ${\underline{u}_0^{N-1}}$ satisfies $\supp({\underline{u}_0^{N-1}})=\mathcal{U}_i(\mathcal{J})$.
    Assume $X$ is the number of the sets $\mathcal{J}$ that 
    $\supp({\underline{u}_0^{N-1}})\neq \mathcal{U}_i(\mathcal{J})$, 
    we have ${\rm A}_{i,\omega_{\min}}^{\rm MPAC} = 2^{|\mathcal{K}_i|} - X$,
    where $X\in[0,2^{|\mathcal{K}_i|}]$.
    The value of $X$ is obtained by assessing the above mentioned 
    criteria \textbf{C1}, \textbf{C2} and \textbf{C3}.
\end{proof}

Recalling (\ref{F_c_piao}), (\ref{F_J}) and the $\mathcal{M}$-construction, 
for every index $f \in \tilde{\mathcal{F}_{\rm c}}\setminus\mathcal{F}_{\mathcal{J}}$, 
$\mathcal{M}(\mathcal{J} \cup \mathcal{F}_\mathcal{J})$ has 
the same property of $|\mathcal{S}_f\setminus\mathcal{S}_i|>1$.
Let us further define set 
$$\mathcal{F}^{*}=\left\{f \in (\mathcal{F}_{\rm c}\cup\mathcal{F}_{\rm p}) \cap [i+1,N-1]  \mid  
\left|\mathcal{S}_f \setminus \mathcal{S}_i\right|>1\right\},$$
where $\mathcal{E}_{\mathcal{J}}^{1}\cup\mathcal{E}_{\mathcal{J}}^{2}\cup\mathcal{E}_{\mathcal{J}}^{3}
\subseteq \mathcal{F}^{*}$.
The following Remark \ref{remark_F_set} reveals that 
if $\mathcal{F}^{*}=\emptyset$, the value of $X$ can be set as $0$
without enumerating the $2^{|\mathcal{K}_i|}$ characterized MWCs. 

\begin{remark}
    \label{remark_F_set}
    Since $\mathcal{E}_{\mathcal{J}}^{1}\cup\mathcal{E}_{\mathcal{J}}^{2}\cup\mathcal{E}_{\mathcal{J}}^{3}
    \subseteq \mathcal{F}^{*}$, if one of the criteria \textbf{C1}, \textbf{C2} and \textbf{C3} is met, 
    there exists at least an index $f$ that satisfies $f\in\mathcal{F}^{*}$.
    If $\mathcal{F}^{*}=\emptyset$, none of the three criteria can be met and $X=0$.
    Note that $\mathcal{F}^{*}=\emptyset$ if and only if 
    $\{(\mathcal{F}_{\rm c}\cup\mathcal{F}_{\rm p}) \cap [i+1,N-1]\}=\emptyset$ or
    $\left|\mathcal{S}_f \setminus \mathcal{S}_i\right|=1$ for all $f\in \{(\mathcal{F}_{\rm c}\cup\mathcal{F}_{\rm p}) \cap [i+1,N-1]\}$.
    The former case implies that every set $\mathcal{U}_i(\mathcal{J})$ is a subset of 
    $\{(\mathcal{I}_{\rm c}\cup\mathcal{I}_{\rm p}) \cap[i,N-1]\}$, which is attainable.
    For the latter, $\mathcal{E}_{\mathcal{J}}^{1}=\mathcal{E}_{\mathcal{J}}^{2}=\mathcal{E}_{\mathcal{J}}^{3}=\emptyset$.
    None of the above cases is consistent with \textbf{C1}, \textbf{C2} and \textbf{C3}. Hence, $X=0$.
\end{remark}

Except the above mentioned case, counting the number of MWCs in coset $\mathcal{C}_i(\mathcal{I})$, 
i.e., ${\rm A}_{i,\omega_{\rm{min}}}^{\rm{MPAC}}$,
can be performed through enumerating all the $2^{|\mathcal{K}_{i}|}$ index sets $\mathcal{U}_i(\mathcal{J})$
and checking the three criteria for related vectors $\underline{u}_0^{N-1}$.
Based on Theorem \ref{theorem_MPAC_number},
we have ${\rm A}_{i,\omega_{\rm{min}}}^{\rm{MPAC}}=2^{|\mathcal{K}_i|} - X$.
An ongoing approach is applied to construct $\mathcal{U}_i(\mathcal{J})$. The values of $u_x$ with $x\in[i,N-1]$ can then be checked such that the values of $X$
and ${\rm A}_{i,\omega_{\rm{min}}}^{\rm{MPAC}}$ can be determined.
This is summarized as in {\bf Algorithm \ref{algorithm_Ad_MPC}},
which evolves from Algorithm 1 of \cite{Rowshan2022FastEnumerationMinimum}.
However, when focusing on the MPAC codes, 
the additional cases should be taken into account.
During the bit-by-bit acquisition of $u_x$ for each $\mathcal{J}\subseteq\mathcal{K}_i$, 
we only record the updates of the shift register states $S$.
Notably, state $S$ is updated only at bits $u_x$ such that
$x\in \mathcal{I}_{\rm c}\cup \mathcal{F}_{\rm c}$.
The ongoing approach initializes $X=0$ and proceeds to enumerate $u_x$ for $x=i$ to $N-1$. According to the reached index of $x$, there are three cases summarized as follows.

\begin{itemize}
    \item When reaching an index $x\in \mathcal{I}_{\rm c}$ or $x\in \mathcal{I}_{\rm p}$,
    if $x\in\mathcal{J}$,
    let $u_x=1$ and partially construct $\mathcal{M}(\mathcal{J} \cup \mathcal{F}_\mathcal{J})$ that is
    denoted as $\tilde{\mathcal{M}}$, using the subroutine \texttt{addToM}($\cdot $)
    that was defined in Algorithm 2 of \cite{Rowshan2022FastEnumerationMinimum}.
    Index $x$ is also needed to add to the partially formed 
    $\mathcal{J}$, which is denoted as $\tilde{\mathcal{J}}$.
    In particular, when $x\in \mathcal{I}_{\rm c}$, the 
    value of $u_x$ can be determined based on (\ref{MPAC_MWC_setting_u}).
    Furthermore, the input bit $s_k$ is determined by (\ref{conv_bit_decision}) 
    and the state $S$ is updated by \texttt{conv1b}($s_k, S, \underline{g}_{0}^{t}$),
    where $k = \digamma_{\mathcal{I}_{\rm c}\cup\mathcal{F}_{\rm c}}(x)$.
    \item When reaching an index $x\in \mathcal{F}_{\rm c}$
    as in \textbf{Case \Rmnum{1}-B}, 
    $u_x$ is determined by the previous convolutional inputs,
    i.e., $[u_x, S] \leftarrow \texttt{conv1b}(0, S, \underline{g}_{0}^{t})$.
    If either criteria \textbf{C1} or \textbf{C2} is satisfied, 
    $X$ is increased by one.
    In particular, if $x\in\mathcal{F}_{\mathcal{J}}$,
    the operations for index $x\in \mathcal{J}$ will be also taken.
    
    It should be mentioned that the subroutine \texttt{addToM}($\cdot$) does not distinguish 
    the indices in sets $\mathcal{J}$ and $\mathcal{F}_{\mathcal{J}}$. 
    \item When reaching an index $x\in \mathcal{F}_{\rm p}$
    as in \textbf{Case \Rmnum{3}}, $u_x$ is frozen as $0$.
    If criterion \textbf{C3} is satisfied, 
    $X$ is also increased by one.
\end{itemize}

With the above mentioned procedure, $X$ can be determined and so can
$A_{i,\omega_{\rm{min}}}^{\rm{MPAC}}$ be obtained.
For every $\mathcal{C}_i(\mathcal{I})$, complexity of the coset-wise enumeration 
is ${\it O}(2^{|\mathcal{K}_i|}\cdot (N-i))$.
Moreover, by introducing a factor $\zeta=\max\{\mathcal{F}^{*}\}$, 
the computational complexity
can be reduced to ${\it O}(2^{|\mathcal{K}_i^{\zeta}|}\cdot (\zeta-i))$.
It follows a similar strategy that was introduced in \cite{Rowshan2022FastEnumerationMinimum}.
Based on {\bf Algorithm \ref{algorithm_Ad_MPC}},
the set $\mathcal{K}_i$ (in line $4$ of {\bf Algorithm \ref{algorithm_Ad_MPC}}) 
and the vector $\underline{u}_i^{N-1}$ (in line $10$ of {\bf Algorithm \ref{algorithm_Ad_MPC}})
degenerate into set $\mathcal{K}_i^{\zeta}=\{j\in\mathcal{K}_i  \mid  j< \zeta \}$ and 
vector $\underline{u}_i^{\zeta}$, respectively.
As a result, 
${\rm A}_{i,\omega_{\min}}^{\rm MPAC}=2^{|\mathcal{K}_i\setminus\mathcal{K}_i^{\zeta}|}
(2^{|\mathcal{K}_i^{\zeta}|}-X)$.

Alternatively, we can also employ an SCL decoder with a sufficiently large decoding output list size $L$, 
which exhibits a computational complexity of ${\it O}(L\cdot N \log_2 N)$ for
enumerating the MWCs \cite{Li2012AdaptiveSuccessiveCancellation}.
But the SCL-based enumeration method is conducted when the all-zero codeword is transmitted at 
a very high SNR, e.g., $20$ dB.
Recalling Theorem \ref{theorem_MPAC_number},
the list size should satisfy $L\geq 2^{|\mathcal{K}_i|}$ for enumerating all
the MWCs in the coset $\mathcal{C}_i(\mathcal{I})$.
Hence, computational complexity of this method is at least
${\it O}(2^{|\mathcal{K}_i|}\cdot N\log_2 N)$.
Since
$2^{|\mathcal{K}_i^{\zeta}|}\cdot (\zeta-i)\ll 2^{|\mathcal{K}_i|}\cdot N\log_2 N$,
the proposed coset-wise enumeration method is much simpler 
than the existing SCL-based enumeration method.

\begin{algorithm}[!thbp]
    \label{algorithm_Ad_MPC}
    \caption{Enumeration of MWCs in coset $\mathcal{C}_i(\mathcal{I})$ of MPAC Code, i.e. $A_{i,\omega_{\rm{min}}}^{\rm{MPAC}}$}
    \KwIn{$i$, $\mathcal{I}_{\rm c}$, $\mathcal{F}_{\rm c}$,  $\mathcal{I}_{\rm p}$, $\mathcal{F}_{\rm p}$, $S$, $\underline{g}_{0}^{t}$;}
    \KwOut{$A_{i,\omega_{\rm{min}}}^{\rm{MPAC}}$;}
    {\bf Initialize} $X=0$ and construct sets $\mathcal{K}_i$ and $\mathcal{F}^{*}$\;
    \If(\tcp*[f]{Remark \ref{remark_F_set}} ){$\mathcal{F}^{*}=\emptyset$}{
        {\bf Return} $A_{i,\omega_{\rm{min}}}^{\rm{MPAC}}=2^{|\mathcal{K}_i|}$\;
    }
    \For{every $\mathcal{J} \subseteq \mathcal{K}_i$}{
        Set $\tilde{\mathcal{J}}, \tilde{\mathcal{M}} \leftarrow \emptyset$\;
        Set $\underline{u}_0^{N-1}=\underline{0}_0^{N-1}, \underline{s}_0^{N_{\rm c}-1} = \underline{0}_0^{N_{\rm c}-1}$ and $S = \underline{0}_0^{t-1}$\;
        \If{$i\in \mathcal{I}_{\rm c}$}{
            Compute $k = \digamma_{\mathcal{I}_{\rm c}\cup\mathcal{F}_{\rm c}}(i)$ and set $s_k = 1$\;
            $[u_i, S] \leftarrow \texttt{conv1b}(s_k, S, \underline{g}_{0}^{t})$\;
        }
        \For{$x = i+1$ \KwTo $N-1$}{
            \uIf{$x \in \mathcal{I}_{\rm c}$ \text{\rm or} $x \in \mathcal{I}_{\rm p}$}{
                \If(\tcp*[f]{for $x\in \mathcal{I}_{\rm c}\cup \mathcal{I}_{\rm p}$}){$x\in \mathcal{J}$}{
                    $\tilde{\mathcal{M}} \leftarrow \texttt{addToM}(i,x, \tilde{\mathcal{J}}, \tilde{\mathcal{M}})$\tcp*{\cite[Alg. 2]{Rowshan2022FastEnumerationMinimum}}
                    $\tilde{\mathcal{J}} \leftarrow \tilde{\mathcal{J}}\cup\{x\}$\;
                }
                \If{$x\in \mathcal{I}_{\rm c}$}{
                    Compute $k = \digamma_{\mathcal{I}_{\rm c}\cup\mathcal{F}_{\rm c}}(x)$\;
                    \eIf{$x\in \mathcal{J}$}
                    {Set $u_x = 1$\;}
                    {
                        \lIf{$x\in \tilde{\mathcal{M}}$}{set $u_x = 1$}
                        \lIf{$x \notin \tilde{\mathcal{M}}$}{set $u_x = 0$}
                    }
                    Compute $s_k=\lambda(u_x, S)$\;
                    Update $S \leftarrow \texttt{conv1b}(s_k, S, \underline{g}_{0}^{t})$\;
                }
            }
            \uElseIf{$x \in \mathcal{F}_{\rm c}$}{
                Compute $k = \digamma_{\mathcal{I}_{\rm c}\cup\mathcal{F}_{\rm c}}(x)$\;
                Compute $[u_x, S] \leftarrow \texttt{conv1b}(0, S, \underline{g}_{0}^{t})$\;
                \eIf{$x \in \mathcal{F}^{*}$}{
                    \If{$u_x=1 \ \text{\rm and}\  x \notin \tilde{\mathcal{M}}$}{
                        $X = X+1$, \textbf{break}\tcp*{C1}
                    }
                    \If{$u_x=0 \ \text{\rm and}\  x \in \tilde{\mathcal{M}}$}{
                        $X = X+1$, \textbf{break}\tcp*{C2}
                    }
                }
                {
                    \If(\tcp*[f]{for $x\in \mathcal{F}_{\mathcal{J}}$}){$u_x=1 $}{
                        $\tilde{\mathcal{M}} \leftarrow \texttt{addToM}(i,x, \tilde{\mathcal{J}}, \tilde{\mathcal{M}})$\;
                        $\tilde{\mathcal{J}} \leftarrow \tilde{\mathcal{J}}\cup\{x\}$\; 
                    }
                }
            }
            \Else(\tcp*[f]{for $x\in \mathcal{F}_{\rm p}$}){
                \If{$x\in\tilde{\mathcal{M}}$}{
                    $X = X+1$, \textbf{break}\tcp*{C3}
                }
            }
        }
    }
    {\bf Return} $A_{i,\omega_{\rm{min}}}^{\rm{MPAC}} = 2^{|\mathcal{K}_i|}-X$\;
\end{algorithm}

\section{AUB-optimal MPAC Codes}
\label{AUBoptimal_MPAC_Codes}
This section designs the AUB-optimal MPAC codes through searching for the optimal convolutional parameters $(N_{\rm c}^*,K_{\rm c}^*)$. 
This design approach shows a partially convolutional transformation can improve the minimum weight distribution for an MPAC code over its prototype PAC code, attributed to utilizing the row combinations of the frozen set more efficiently. 
The AUB-optimal MPAC code outperforms a PAC code and a CRC-polar code in terms of both MWD and decoding performance.

\subsection{Design of AUB-Optimal MPAC Codes}
With the proposed coset-wise analysis of the MWD for MPAC codes, 
the minimum distance $d_{\rm min}$ and the number of MWCs 
${\rm A}_{d_{\min}}$ can be determined.
Furthermore, eq. (\ref{AUB}) implies that both maximizing $d_{\rm min}$
and minimizing ${\rm A}_{d_{\min}}$ can optimize the AUB $P_e$.
The asymptotic ML decoding performance of different MPAC codes that are specified by 
their convolutional parameters $(N_{\rm c},K_{\rm c})$ can be evaluated by their corresponding AUBs $P_e$.
It guides the design of MPAC codes,
especially in determining $(N_{\rm c},K_{\rm c})$ in optimizing the code's AUB performance.
We call them the \textit{AUB-optimal MPAC codes}.

Given $N$ and $K$, let $(N_{\rm c}^*,K_{\rm c}^*)$ denote the pair that optimizes the code's AUB performance.
Considering the parameter constraint of (\ref{parameter_constraint}), 
we have
\begin{equation}
    \label{search_space}
    \begin{matrix}
        0\le N_{\rm c}-K_{\rm c}\le N-K, \\
        0< K_{\rm c}\le K, \\
        0< N_{\rm c}\le N.
    \end{matrix}
\end{equation}
The above inequalities imply that
$K_{\rm c}\in[1, K]$ and  $N_{\rm c}\in[K_{\rm c}, N-K+K_{\rm c}]$.
There are $K(N-K+1)$ pairs of $(N_{\rm c},K_{\rm c})$.
With the coset-wise analysis of MWD, 
this parameter search becomes practical.
Since the two rate profilings and the choice of $(N_{\rm c},K_{\rm c})$ jointly determine 
$d_{\rm min}$,
maximizing $d_{\rm min}$ can be realized by choosing specific pairs of 
$(N_{\rm c},K_{\rm c})$.
This helps rationalize the otherwise exhaustive search process.

\begin{figure}[!htbp]
	\centering
	
    \includegraphics[scale=0.60]{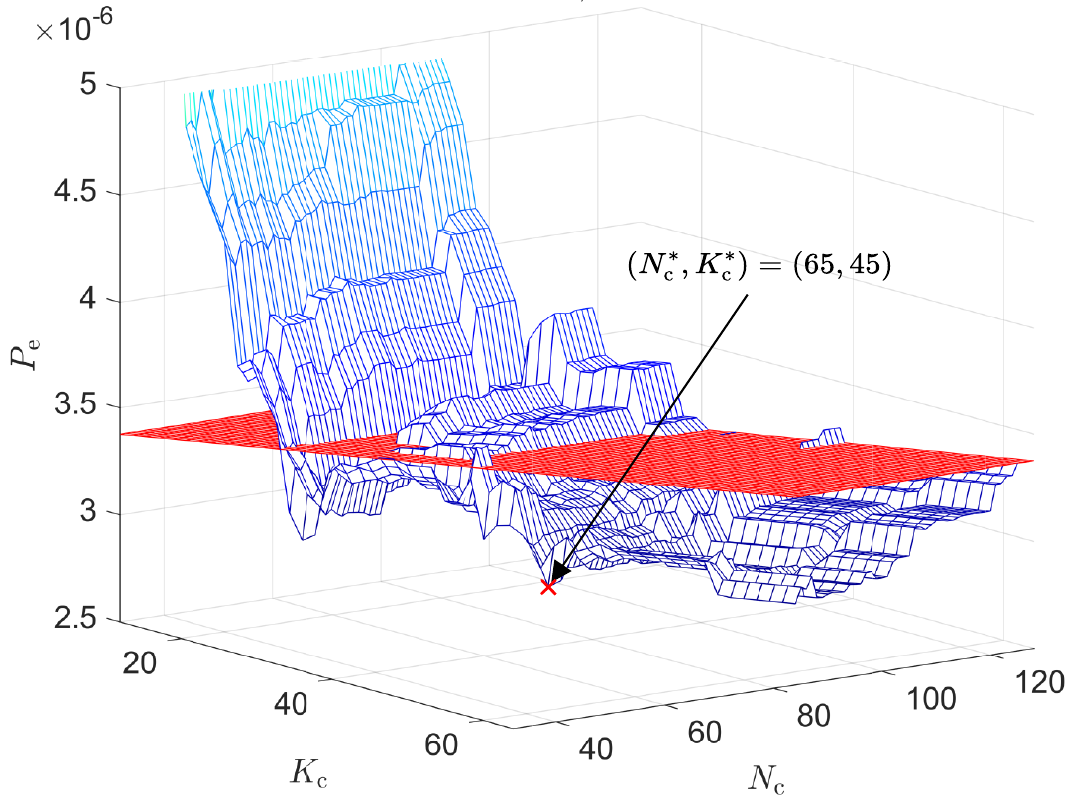}
	\caption{AUB performances of the MPAC codes and PAC codes versus parameters $(N_{\rm c},K_{\rm c})$, 
    where $N = 128$, $K = 64$ and $E_{\rm b} / N_0=3.5$ dB.}
	\label{fig_Nc_Kc_search}
\end{figure}

Fig. \ref{fig_Nc_Kc_search} shows how the AUB performances $P_e$ of MPAC codes are affected
by $K_{\rm c}$ and $N_{\rm c}$ under the SNR per information bit of $E_{\rm b} / N_0=3.5$ dB.
We consider the MPAC codes with length $N=128$, dimension $K=64$ and the 
convolutional generator polynomial coefficients of $(1,0,1,1,0,1,1)$.
The rate profiling is conducted by GA at an SNR of $0$ dB.
We compare the AUB performances of MPAC codes with the prototype $(128, 64)$ PAC code of \cite{Arikan2019sequentialdecodingchannel}.
The PAC code is constructed by the RM rate profiling.
Its MWD is $d_{\rm min}=16$ and ${\rm A}_{d_{\rm min}}=3120$.
The AUB performances are measured versus parameters $(N_{\rm c},K_{\rm c})$, 
yielding the blue and the red meshes that correspond to the MPAC codes and the PAC code,
respectively.
In particular, the AUB-optimal MPAC code
with parameter $(N_{\rm c}^*,K_{\rm c}^*)=(65,45)$
is marked by a red `$\times$' symbol.
The code has $d_{\rm min}=16$ and ${\rm A}_{d_{\rm min}}=2633$.
It can be seen that there exist numerous $(N_{\rm c},K_{\rm c})$ 
that can enable an MPAC code to outperform the prototype PAC code.
This is due to a better weight distribution is produced by 
the partially convolutional transform in the MPAC codes.
Furthermore, Fig. \ref{fig_Nc_Kc_search} also suggests that 
a larger $K_{\rm c}$ is more likely to yield a better weight distribution for the MPAC codes.
It helps maximize $d_{\rm min}$ and 
yields a greater MWC number deficit $X$.
With a large $K_{\rm c}$, reducing $N_{\rm c}$ within the
scope that keeps $d_{\rm min}$ being maximized generally achieves a better AUB performance. 
This indicates that assigning elements from set $\mathcal{F}_{\rm c}$ to set $\mathcal{F}_{\rm p}$ can lead to a better MWD for the MPAC codes. 
An extreme case to the opposite is when $K_{\rm c}=0$,
the MPAC codes become polar codes with no improvement on its weight distribution.

\subsection{Simulation Results}
\label{sec4_Simulation_Results}

This subsection presents the MWD results and performances of the AUB-optimal MPAC codes.
The AUB-optimal MPAC codes are also compared with the relevant coding schemes, 
especially the MPAC codes designed in Section \ref{MPAC_HFSC}.
Note that the parameters of the relevant coding schemes that are applied in this subsection
are the same as those in Section \ref{optimize_HFSC_performance}.

\begin{table}[!ht]
	\centering
    \caption{The MWD of AUB-optimal MPAC codes, PAC codes, ensemble average and CRC-polar codes 
    with different code parameters}
	\label{table_chap4_AUB_result}
	\setlength{\tabcolsep}{2.0mm} \renewcommand{\arraystretch}{1.25}
	\begin{tabular}{|c|c|c|c|c|c|c|c|c|c|c|}
		\hline
		\multirow{2}{*}{$N$} & \multirow{2}{*}{$K$} & \multicolumn{2}{c|}{PAC} & \multicolumn{3}{c|}{MPAC} & \multicolumn{2}{c|}{Ensemble Average} & \multicolumn{2}{c|}{CRC-polar} \\ \cline{3-11}
		&  & $d_{\min}$ & ${\rm A}_{d_{\min}}$ & $(N_{\rm c}^*,K_{\rm c}^*)$ & $d_{\min}$ & ${\rm A}_{d_{\min}}$ & $d_{\min}$ & ${\rm A}_{d_{\min}}$ & $d_{\min}$ & ${\rm A}_{d_{\min}}$\\ \hline
		\multirow{3}{*}{128} & 29 & 32 & 820 & (61,29) & 32 & 436 & 32 & 422.1 & 24 & 53 \\ \cline{2-11}
		& 64 & 16 & 3120 & (65,45) & 16 & 2633 & 16 & 2766.9 & 12 & 224 \\ \cline{2-11}
		& 99 & 8 & 16080 & (89,71) & 8 & 15264 & 8 & 15936.0 & 6 & 152 \\ \hline
		\multirow{2}{*}{256} & 93 & 32 & 5092 & (145,86) & 32 & 2788 & 32 & 2766.9 & 16 & 6 \\ \cline{2-11}
		& 163 & 16 & 18896 & (169,132) & 16 & 15622 & 16 & 15936.3 & 8 & 4 \\ \hline
		\multirow{3}{*}{512} & 130 & 64 & 17092 & (252,110) & 64 & 3466 & 64 & 2766.9 & 32 & 256 \\ \cline{2-11}
		& 256 & 32 & 35900 & (342,222) & 32 & 17108 & 32 & 15936.3 & 16 & 966 \\ \cline{2-11}
		& 382 & 16 & 128704 & (314,216) & 16 & 104209 & 16 & 101280.0 & 8 & 2307 \\ \hline
	\end{tabular}
\end{table}

Table \ref{table_chap4_AUB_result} provides the MWD of PAC codes, the AUB-optimal MPAC codes, the ensemble average of pre-transformed polar codes \cite{Li2021WeightSpectrumPrea} and CRC-polar codes. The average MWD of pre-transformed polar codes is analyzed by the probabilistic method that calculates the weight spectrum of the pre-transformed polar codes with randomly generated pre-transformed matrix and the RM rate profiling.
The detailed procedure of this method can be found in \cite{Li2021WeightSpectrumPrea}.
MPAC codes are constructed by the proposed iRMP construction, while PAC codes are constructed by the RM rate profiling.
The convolutional generator polynomial coefficients $(1,0,1,1,0,1,1)$ are employed.
The CRC-polar codes are constructed by the 5G reliability sequence.

It can be seen that for a wide range of code parameters,
the AUB-optimal MPAC codes yield a better MWD than the respective PAC codes and CRC-polar codes.
In particular, the AUB-optimal MPAC codes yield the same minimum distance $d_{\min}$ as 
the PAC codes, but they contain less MWCs.
Meanwhile, it can also be observed that the minimum distance of the AUB-optimal MPAC codes are greater than
that of the CRC-polar codes. 
Furthermore, the AUB-optimal MPAC codes yield a close ${\rm A}_{d_{\min}}$ compared to the average pre-transformed polar codes.
They demonstrate that the AUB-optimal MPAC codes have the MWD advantage over both the PAC codes and the CRC-polar codes.

With code parameters of $N$ and $K$ concerned in Table \ref{table_chap4_AUB_result}, MPAC codes have the same information set $\mathcal{I}$ as the corresponding PAC codes. Their main difference is reflected in pre-transformed design. By defining $\underline{d}$ as bits before polar transformation, the pre-transformed matrix $\mathbf{G}_{\mathrm{pre}}$ can be utilized to represent the constraint relationship between bits in $\underline{d}_{0}^{N-1}$. 
Note that $\underline{c}_{0}^{N-1}=\underline{d}_{0}^{N-1} \mathbf{G}_{\mathrm{pre}}\mathbf{G}_{\mathrm{p}}$. 
In particular, for PAC codes, we have
$\mathbf{G}_{\mathrm{pre}}=\mathbf{G}_{\mathrm{c}}$. 
For MPAC codes, the submatrix of $\mathbf{G}_{\mathrm{pre}}$ with rows and columns selected in $\mathcal{P}$ is $\mathbf{G}'_{\mathrm{c}}$. $\mathbf{G}_{\mathrm{pre}}$ is further obtained with diagonal elements set to $1$ and other elements set to $0$.

\begin{figure}[htb]
	\centering
	\includegraphics[scale=0.43]{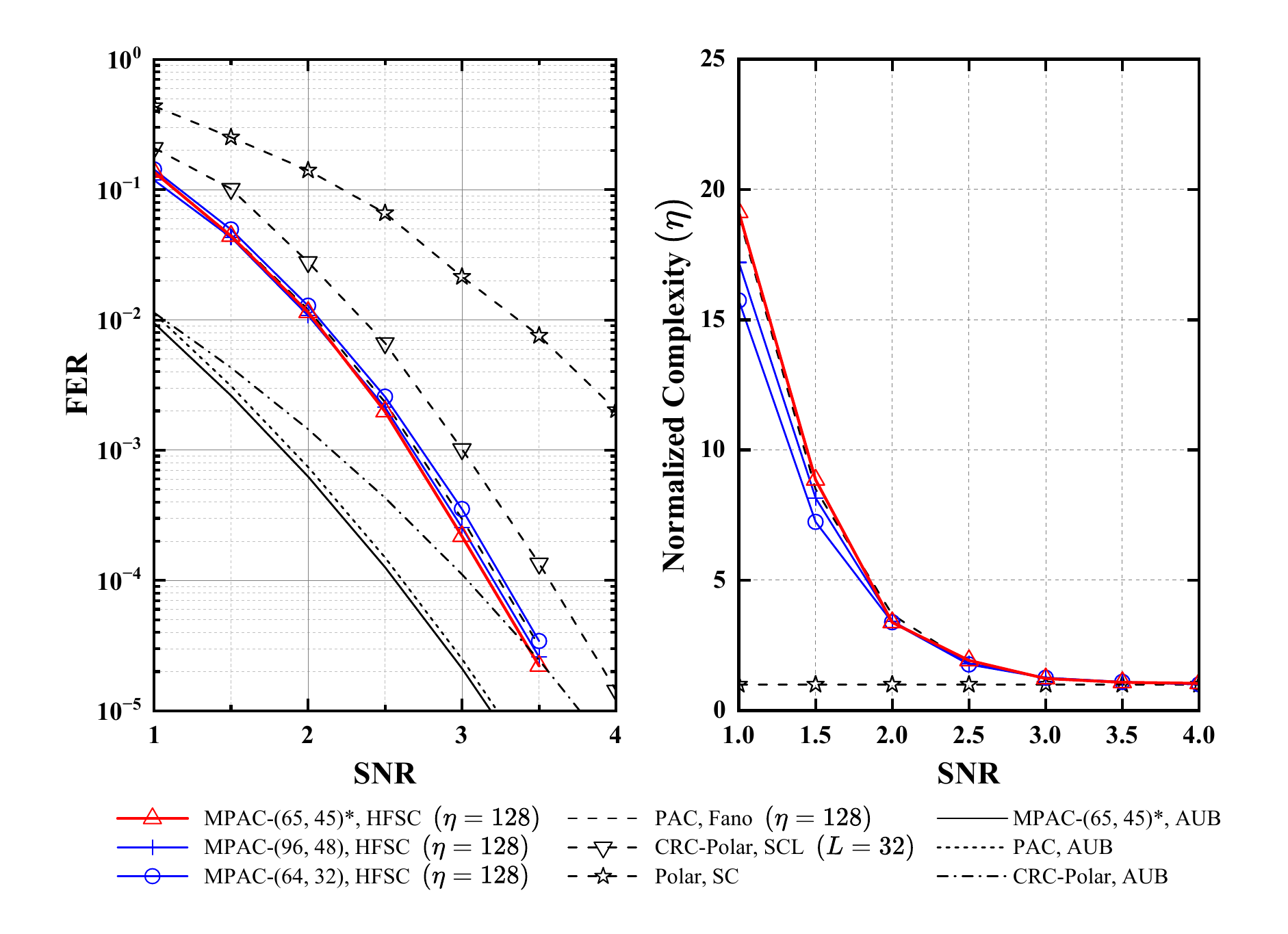}
	\vspace{-0.1cm}
    \caption{Performance comparison of the AUB-optimal MPAC codes and other relevant codes, where $N = 128$ and $K = 64$.}
	\label{fig_simulation_128_64}
\end{figure}

Therefore, the $\mathbf{G}_{\mathrm{pre}}$ of an MPAC code is sparser than that of a PAC code and generally not a Toeplitz matrix.
Consequently, some bits of an MPAC code in $\underline{d}_{\mathcal{I}_{\rm{c}}}$ can influence other bits with a more distant index in $\underline{d}_{\mathcal{F}_{\rm{c}}}$ than the case in PAC code.
For low rate polar-like codes, the continuous length of the frozen set is large. With short convolutional generator polynomials employed, PAC codes can only utilize a portion of $\underline{d}_{\mathcal{F}_{\rm{c}}}$ to reduce the number of MWCs sufficiently.
However, the MPAC codes can utilize the sparser pre-transformed matrix to involve more rows in $\mathcal{F}_{\rm{c}}$ in row combinations. This results in a better effect on the MWC number deficit $X$ increasing especially for low code rates.
As a result, the MPAC codes not only inherit the advantage of PAC codes, but also incorporate the row combinations of the frozen set. After selecting the convolutional parameters, this leads to a better MWD for the AUB-optimal MPAC codes than the PAC codes and the CRC-polar codes.

Note that as the code rate increases, the MWD improvement margin of MPAC codes becomes less significant.
This is because in such situation, some MWCs in the specific cosets cannot be reduced.
The proportion of these irreducible MWCs increases with the code rate. 
On the other hand, higher rate restricts the size of $\mathcal{F}_{\rm p}$ for MPAC codes, which leads to less improvement of MWD in comparison with PAC codes.

\begin{figure}[htb]
	\centering
	\includegraphics[scale=0.47]{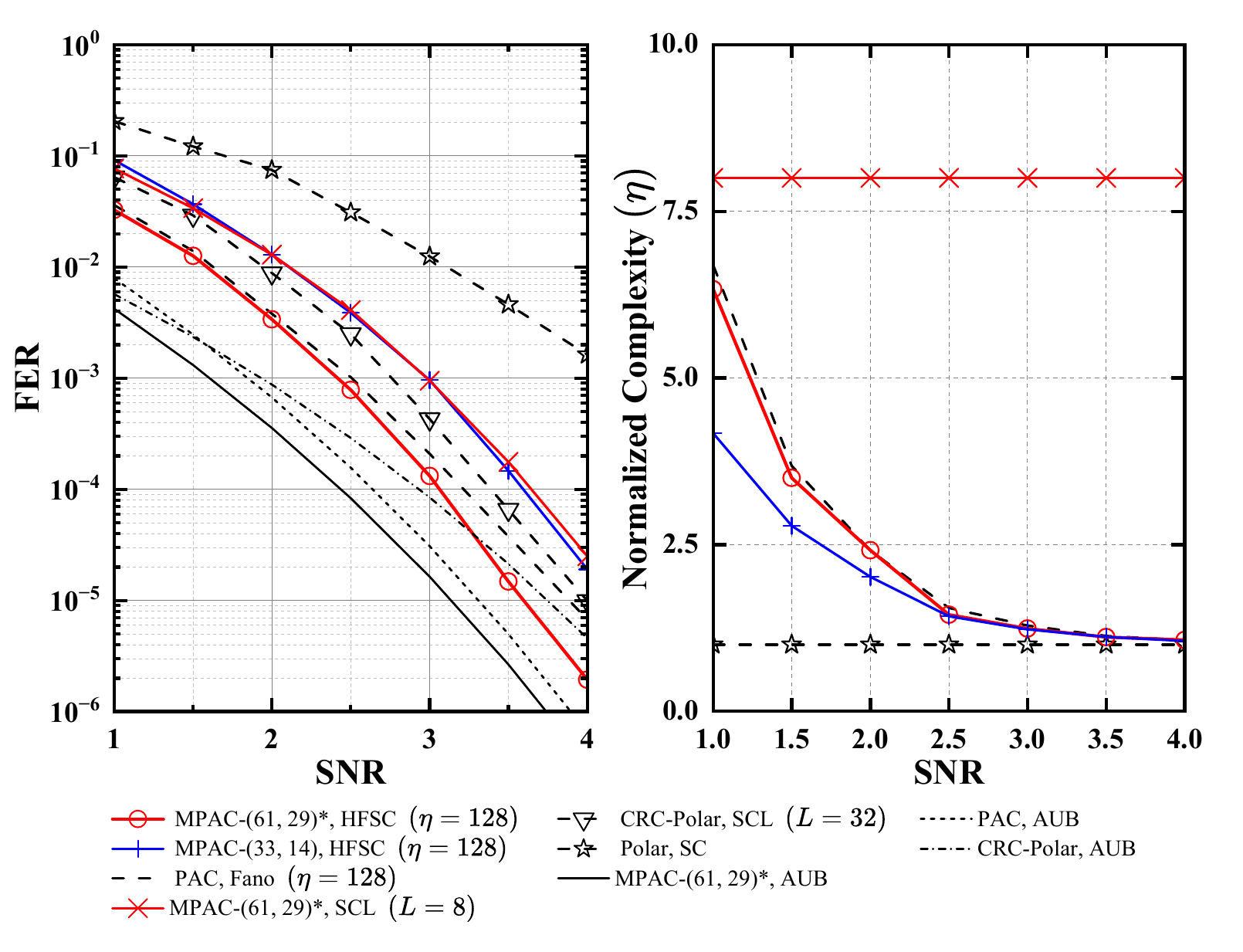}
    \vspace{-0.1cm}
	\caption{Performance comparison of the AUB-optimal MPAC codes and other relevant codes, where $N = 128$ and $K = 29$.}
    \label{fig_simulation_128_29}
\end{figure}

Finally, Fig. \ref{fig_simulation_128_64} shows the HFSC decoding FER, normalized complexity and AUB of 
the optimized MPAC code with $N=128$ and $K=64$. 
A decoding computation threshold of $\eta=128$ is applied. 
The decoding FER and complexity performances of different coding schemes are also provided,
including the MPAC-$(96,48)$ and the MPAC-$(64,32)$ codes.
Note that the MWD of the MPAC-$(96,48)$ code and the MPAC-$(64,32)$ code are 
$\{d_{\min}=16, {\rm A}_{d_{\min}}=2900\}$ and 
$\{d_{\min}=16, {\rm A}_{d_{\min}}=3121\}$, respectively. 
It can be seen that the AUB-optimal MPAC code indicated by $*$ as MPAC-$(65,45)^*$ can slightly outperform 
the MPAC-$(96,48)$ code and the MPAC-$(64,32)$ code with an increased complexity.
Fig. \ref{fig_simulation_128_29} shows the same coding schemes with $N=128$ and $K=29$.
It can be seen that the AUB-optimal MPAC code with HFSC decoding
outperforms the MPAC-$(33,14)$ code, the PAC code and the CRC-polar code
in both the decoding FER and the AUB.

\section{Conclusion}
\label{sec7}
This paper has proposed the MPAC codes and their HFSC decoding. 
Realized by two rate profiling steps with convolutional parameters,
a subset of the information bits first undergo the convolutional transform. 
The convolutional output then together with the remaining information bits 
undergo the polar transform. 
Consequently, the HFSC decoding deploys the Fano decoding and SC decoding 
to recover the information bits that have undergone the convolutional transform and 
those that have not, respectively. 
The MWD analysis of the MPAC codes has been provided in the coset-wise framework, 
which can be further utilized to optimize the design of MPAC codes, 
especially in selecting the convolutional parameters. 
Consequently, the AUB-optimal MPAC codes can be designed, 
achieving a better MWD than the PAC codes and CRC-polar codes. 
The implication of the MWD improvement is examined by effectively exploiting the row combinations of the frozen set, which originates from the sparser pre-transformed matrix of MPAC codes. 
Simulation results of the proposed schemes and the relevant schemes 
have been presented to verify the decoding performance and complexity advantages 
of the proposed MPAC coding schemes.

\section*{Acknowledgement}
This work was sponsored by the National Natural Science
Foundation of China (NSFC) with project ID 62471503 
and Guangdong National Science Foundation (GDNSF) with project ID 2024A1515010213.

\bibliographystyle{IEEEtran}
\bibliography{channel_coding}

\end{document}